\documentclass[jgr, galley]{agutex}

\usepackage{dcolumn}
\usepackage{bm}

\usepackage{lineno}
\usepackage{graphicx}

\authorrunninghead{URITSKY ET AL.}

\titlerunninghead{Hot flow anomalies at Mercury}

\authoraddr{V. M. Uritsky, Catholic University of America at NASA Goddard Space Flight Center,
Code 671, Greenbelt, MD 20771, USA. (vadim.uritsky@nasa.gov)}

\linenumbers*[1]
\begin{document}

\setkeys{Gin}{draft=false}

\title{Active current sheets and hot flow anomalies in Mercury's bow shock}

\author{ V. M. Uritsky\altaffilmark{1, 2}, J. A. Slavin\altaffilmark{3}, S. A. Boardsen\altaffilmark{1, 4}, 
T. Sundberg\altaffilmark{1, 5}, J. M. Raines\altaffilmark{3},  D. J. Gershman \altaffilmark{3}, G. Collinson\altaffilmark{1},  D. Sibeck\altaffilmark{1}, G. V. Khazanov\altaffilmark{1}, B. J. Anderson\altaffilmark{6}, and H. Korth\altaffilmark{6} }

\altaffiltext{1}
{NASA Goddard Space Flight Center, Greenbelt, MD, USA}
\altaffiltext{2}
{Catholic University of America, Washington, DC, USA}
\altaffiltext{3}
{University of Michigan, Ann Arbor, MI, USA}
\altaffiltext{4}
{University of Maryland, Baltimore County, Baltimore, MD, USA}
\altaffiltext{5}
{Center for Space Physics, Boston University, Boston, MA, USA}
\altaffiltext{6}
{John Hopkins University Applied Physics Laboratory, Laurel, MD, USA}
\begin{abstract}

Hot flow anomalies (HFAs) represent a subset of solar wind discontinuities interacting with collisionless bow shocks. They are typically formed when the normal component of motional (convective) electric field points toward the embedded current sheet on at least one of its sides. The core region of an HFA contains hot and highly deflected ion flows and rather low and turbulent magnetic field. In this paper, we report first observations of HFA-like events at Mercury identified over a course of two planetary years. Using data from the orbital phase of the MErcury Surface, Space ENvironment, GEochemistry, and Ranging (MESSENGER) mission, we identify a representative ensemble of active current sheets magnetically connected to Mercury's bow shock. We show that some of these events exhibit unambiguous magnetic and particle signatures of HFAs similar to those observed earlier at other planets, and present their key physical characteristics. Our analysis suggests that Mercury's bow shock does not only mediate the flow of supersonic solar wind plasma but also provides conditions for local particle acceleration and heating as predicted by previous numerical simulations. Together with earlier observations of HFA activity at Earth, Venus and Saturn, our results confirm that hot flow anomalies are a common property of planetary bow shocks, and show that the characteristic size of these events is of the order of one planetary radius.

\end{abstract}


\begin{article}

\section{Introduction}

The MErcury Surface, Space ENvironment, GEochemistry, and Ranging (MESSENGER) mission \citep{solomon01} provides a deep insight into the structure and dynamics of various plasma regions surrounding the planet. The data reveal a rather active plasma environment which is in many respects unique within our solar system. The magnetosphere of Mercury has been intensively studied in the context of tail and magnetopause reconnection, magnetic flux transport, ULF waves and oscillations, propagating dipolarization fronts, and other phenomena (see e.g \cite{anderson08, slavin08, slavin09, slavin09b, boardsen09, slavin10a, sundberg10, sundberg12}). It has been shown that the local interplanetary medium surrounding the planet exhibits turbulent variability over both magnetohydrodynamic and kinetic plasma scales \citep{korth10, uritsky11}. This broadband variability should have a significant impact on the Hermean magnetosphere and its response to the solar wind driver. 

The present paper focuses on dynamic discontinuities in the Hermean foreshock associated with kinetically active current sheets, focusing on the phenomenon of the hot flow anomaly (HFA) \citep{schwartz85, thomsen86}. Using the first 180 days of MESSENGER operation after its orbital insertion, we identify a set of interplanetary current sheets magnetically connected to the bow shock. We show that some of these current sheets exhibit well-defined HFA signatures. We also investigate the influence of HFAs on magnetic field variability in the adjacent plasma regions and find evidence for HFA-triggered ultra low frequency (ULF) waves in quasi-parallel shock configurations. 

The paper is organized as follows. In Section 2, we present a concise review of HFA observations in planetary bow shocks and summarize magnetic and kinetic signatures of HFA events.  Section 3 describes the methodology of our study. Section 4 reports case studies of ten HFA events and several examples of active helio current sheets not interacting with Mercury's bow shock. Section 5 reports statistical properties of the observed events including their location, geometry, duration, relative occurrence rates, and other characteristics. Section 6 summarizes the obtained results.

\section{Hot flow anomalies in planetary foreshocks}

Collisionless planetary bow shocks do not only mediate the flow of supersonic plasma but also provide conditions for particle acceleration and heating. They can energize, decelerate, and deflect solar wind plasma allowing it to flow through the magnetosheath and around the magnetosphere \citep{omidi07}. 
For certain shock geometries, the inflowing solar wind plasma can partly return to the upstream region. The interaction between this counterstreaming particle population and the inflowing plasma naturally leads to various plasma instabilities and waves which may effectively energize ions and electrons \citep{eastwood05}. 

For steady solar wind conditions, the large-scale phenomenology of the foreshock can be organized by the angle $\theta_{B:BS}$ between the upstream magnetic field and the bow shock normal, with the quasi-parallel geometry ($\theta_{B:BS} < 40^o$) producing the most extended and dynamic foreshock system populated by backstreaming ions. Changes in the interplanetary magnetic field (IMF) direction give rise to a variety of small-scale and/or transient foreshock phenomena.

HFAs represent a subset of solar wind discontinuities (rotational or tangential)  interacting with the bow shock \citep{schwartz85, thomsen86, schwartz00, billingham11}. They are formed when the normal component of motional electric field points toward the embedded current sheet on at least one of its sides \citep{schwartz00}. The core regions of HFAs typically contain hot and highly deflected ion flows often described by nearly Maxwellian and isotropic particle distributions, and rather low and turbulent magnetic fields. The direction of bulk plasma flows in HFAs can differ significantly from that of the ambient solar wind plasma. 

The first observations of HFAs near Earth were reported by \citet{schwartz85} and \citet{thomsen86} based on the data from AMPTE and ISEE and missions. Subsequent studies have shown that HFA events appear systematically in the terrestrial foreshock, with the average occurrence rate of about 3 events per day \citep{schwartz00}. They typically last for a few minutes and have spatial scales of the order of one Earth's radius. HFAs can generate considerable perturbations of the dynamic pressure in the upstream solar wind \citep{sibeck99, eastwood08} and induce significant magnetospheric response, including displacement the nominal magnetopause position accompanied by auroral brightening \citep{sibeck99}, riddling of peripheral boundary layers \citep{savin12}, transient ULF geomagnetic pulsations \citep{eastwood11}, and other effects.

HFA activity has also been detected at several other planets. Mars Global Surveyor observed a HFA-like hot diamagnetic cavity upstream of the Martian foreshock \citep{oieroset01}. More recently, HFAs were found at Saturn's bow shock based on Cassini data \citep{masters08, masters09}. Magnetic signatures of HFA events at Venus were first reported by \citet{slavin09a} using MESSENGER magnetometer data. The presence of HFAs at Venus was later confirmed by magnetic, electron and ion observations from Venus Express \citep{collinson12}. Whilst HFAs have been found throughout the solar system, none have been observed at Mercury until now. 

\section{Data and methods}

\subsection{MESSENGER's orbit and data}

We investigated the first 180 days of the MESSENGER orbital operations (24 March - 19 Sep, 2011) corresponding to two Mercury years. During this time, MESSENGER followed a highly elliptical orbit (periapsis $\sim$200 km, apoapsis $\sim$15,193 km) enabling observations of a significant part of the Mercury's foreshock both in the dawn and dusk sectors. The orbit was inclined $82.5^{\circ}$ to the equator. 

The magnetic field data were obtained from the MAG magnetometer \citep{anderson07}. The three magnetic field components were measured with a three-axis, ring-core fluxgate detector at a typical sampling period $\Delta t =$50 ms. MAG data were used to locate interplanetary active current sheets connected with the Hermean bow shock, characterize their dynamics and geometry in terms of previous HFA studies, and identify likely instances of hot flow anomaly events.

For some of the events, we also used the data from the Fast Imaging Plasma Spectrometer(FIPS, \citet{andrews07}). The ion plasma instrument FIPS onboard MESSENGER measures ions in the energy per charge range of 50 eV/q to 20 keV/q and in the mass per charge range of 1 amu/q to 40 amu/q. The FIPS spectral data used in this study were averaged over the angular field of view of 1.4 $\Omega$, of which 0.4 $\Omega$ is obscured by the solar array panels, spacecraft body, and heat shield \citep{raines11} so angular distribution information is not yet available for this data type. The sunshade mounted on MESSENGER's spacecraft body nominally blocks observation of the centroid of the solar wind ion velocity space distributions by FIPS.  The most limiting factor in identifying HFA signatures in FIPS observations, however, is the time resolution.  The time required for FIPS to complete an energy-scan is 64 s or 8 s, depending on the instrument operational mode.  

Since the FIPS count rates are a function of plasma drift velocity, temperature, and orientation, this data should be interpreted with caution.  We used FIPS observations to look for the presence of ions with energies atypical of the expected solar wind, leaving a more detailed analysis for future studies.

The Mercury Solar Orbital (MSO) coordinate system is used for all vector quantities, with $X_{MSO}$ directed from the center of the planet toward the Sun, $Z_{MSO}$ being perpendicular to Mercury's orbital plane and pointing toward the north celestial pole, and $Y_{MSO}$ completing the right-handed system. 

\subsection{Initial event detection}

Since MESSENGER FIPS is not a reliable source of information on transient localized plasma processes at Mercury, our event detection was based on MAG data. The core regions of HFAs usually contain intervals of considerable magnetic depression due to the high particle pressure exerted by the hot ions. We used this signature as the starting point of our search for Mercury's HFAs, followed by the analysis of more subtle features including current sheet geometry and detailed B-field variation, reinforced by the analysis of particle data whenever these data were available.

\begin{figure}
\begin{center}
\noindent\includegraphics*[width=15 cm]{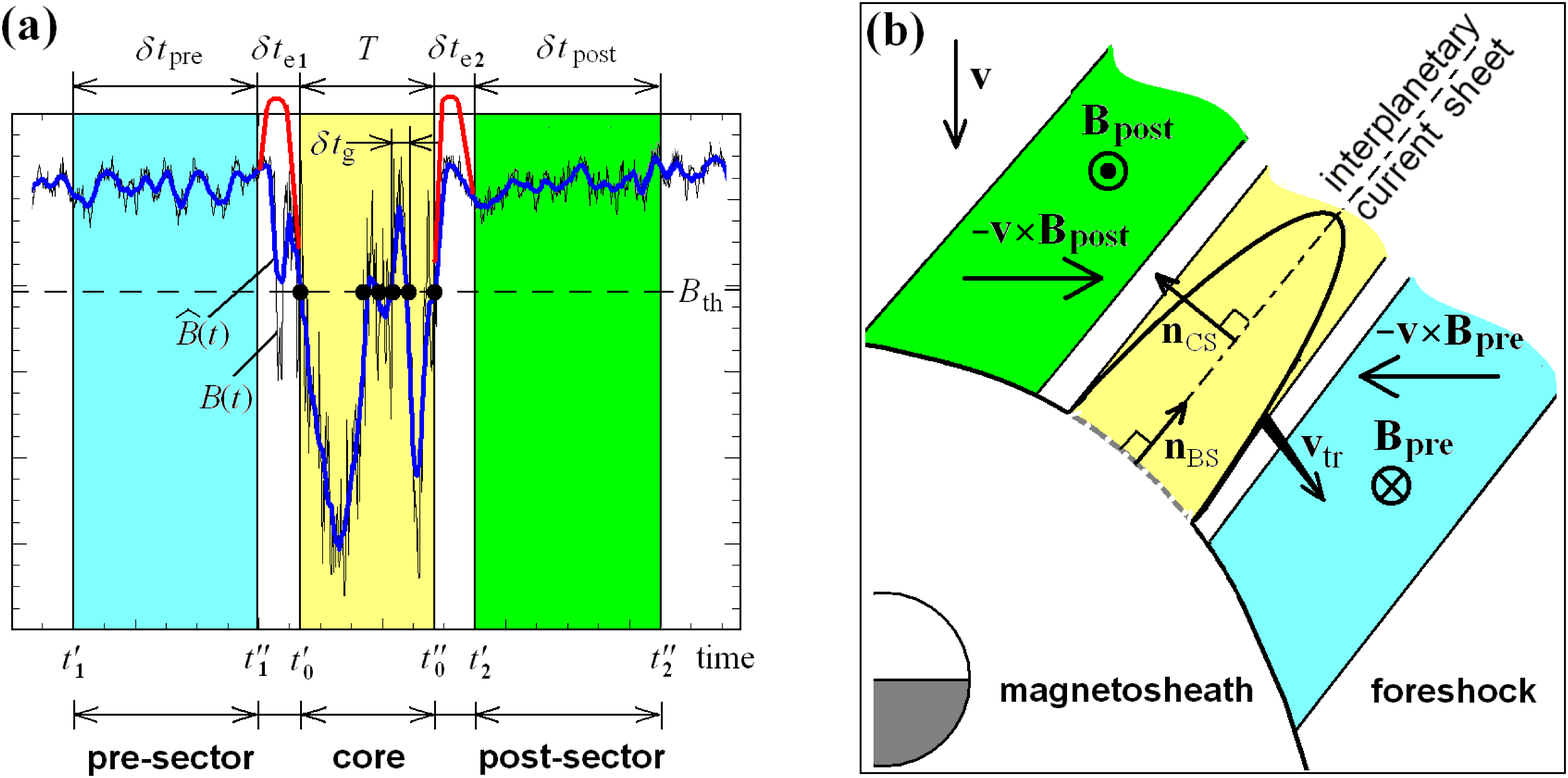}
\end{center}
\caption{\label{fig1} (a): Illustration of the threshold-based detection of an HFA event in the Hermean foreshock. (b): Spatial structure of the event. See text for details.}
\end{figure}

Fig. \ref{fig1} illustrates our event detection criteria and the associated time intervals. The hot core region of the event shown with a yellow rectangle is embedded in a cooler plasma medium which is encountered before and after the event. These encounters  are labeled as the pre-sector (blue rectangle) and the post-sector (green rectangle), correspondingly. The red magnetic shoulders surrounding the core region are indicative of terrestrial HFA events and are a signature of plasma compression caused to an expansion of the core HFA region. Such magnetic shoulders can be seen in a proto-form in some of the Mercury events reported here, although we were unable to find any Mercury events with fully developed compression edges. 

The intervals of the depressed magnetic field were identified using the smoothed magnetic field $\widehat{B}$
\begin{equation}\label{eq1}
\widehat{B}(t_i) = \frac{1}{w} \sum_{k=i-w/2}^{i+w/2-1} B(t_k)\\
\end{equation}
subjected to the threshold condition
\begin{equation}\label{eq1a}
\widehat{B}(t) < B_{th},
\end{equation}
where $w$ is the size of the moving window in time step units. In the presence of fast fluctuations (time scale $\tau \ll w  \Delta t$), the smoothed signal $\widehat{B}(t)$ enables more reliable detection of threshold crossings compared to the raw magnetic signal. Low-frequency fluctuations with $\tau \geq w  \Delta t$ require additional attention since they can cause $\widehat{B}$ to cross the threshold more than once during the same event. Such fluctuations are commonly observed in the core HFA region where their amplitude can reach the field strength in the surrounding plasma (see e.g. \citet{paschmann88}). To properly attribute multiple $B_{th}$ crossings due to such fluctuations to a single magnetic depression event, we merged together transient magnetic decreases separated by a time gap $\delta t_g$ of less than 30 seconds, with the initial smoothing interval of 1 second ($w = 20$).

After a depressed B-field event is detected, we identified three main time intervals corresponding to the observations in the core region of the event ($t \in [ t'_0, t''_0 ]$) as well as in its pre-sector ($t \in [ t'_1, t''_1 ]$) and the post-sector ($t \in [ t'_2, t''_2 ]$). The boundaries of the three intervals were calculated as follows (see also Fig. \ref{fig1}):
\begin{eqnarray}\label{eq2} 
t'_0 & = & \mbox{min} \{ t | \widehat{B}(t) < B_{th} \} \\
t''_0 & = & \mbox{max} \{ t | \widehat{B}(t) < B_{th} \} \\
t'_1 & = & t'_0 - \delta t_{e1} - \delta t_{pre} \\
t''_1 & = & t'_0 - \delta t_{e1} \\
t'_2 & = & t''_0 + \delta t_{e2} \\
t''_2 & = & t''_0 + \delta t_{e2} + \delta t_{post}.
\end{eqnarray}
Here, $B_{th}$ is the detection threshold, $\delta t_{pre}$ ($\delta t_{post}$) is the duration of the pre (post) sector,  $\delta t_{e1}$ and $\delta t_{e2}$ are the sizes of the edge regions flanking the core region from either side \citep{schwartz95}. The default values $\delta t_{pre}= \delta t_{post} = 30$ s and $\delta t_{e1} = \delta t_{e2} = 10$ s were used for the automatic identification of the three regions in the entire set of the detected events. The region boundaries of the events showing clear HFA signatures were then readjusted manually taking into account a particular shape of the B-field variation. The locations of the events were evaluated based on the average MSO position of the core region; the event duration $T$ was calculated from the time boundaries of this region:

\begin{equation}\label{eq5}
T = t''_0 - t'_0
\end{equation}

\subsection{Magnetic geometry}

For each magnetic field depression event, we computed the current sheet normal as the cross product between magnetic field before and after the event:

\begin{equation}\label{eq3}
\mathbf{n}_{CS} = \pm \mathbf{B}_1 \times \mathbf{B}_2,
\end{equation}
in which $\mathbf{B}_1$ and $\mathbf{B}_2$ are the average magnetic field vectors in the pre- and post- sectors, correspondingly. The sign ambiguity was resolved by requiring $\mathbf{n}_{CS} \cdot \mathbf{V}_{SW} < 0$, as appropriate for HFA studies \citep{schwartz00}. The solar wind velocity $\mathbf{V}_{SW}$ was assumed to be strictly anti-sunward and hence parallel to the $X_{MSO}$ axis. We also determined the shortest (projection) distance $d_{BS}$ from the core region of the event to the model bow shock surface describing the average position of Mercury's bow shock for a solar wind fast mode Mach number $\sim 3$ \citep{slavin09}. The same model was used to calculate the local bow shock normal $\mathbf{n}_{BS}$ attached to the projection point. 

As stated above, a key HFA formation condition is that the solar wind convection (motional) electric field $\mathbf{E} = - \mathbf{V}_{SW} \times \mathbf{B}$ points into the underlying discontinuity on at least one side (see Fig. \ref{fig1}, right panel). This condition was verified by computing the angles $\theta_{E1:CS}$ and $\theta_{E2:CS}$ between the current sheet normal and the average electric field in respectively pre- and post- sectors. We also computed the angle $\theta_{B1:B2}$ between the magnetic vectors $\mathbf{B}_1$ and $\mathbf{B}_2$, the angles $\theta_{B1:BS}$ and $\theta_{B2:BS}$ created by these vectors with the local bow shock normal $\mathbf{n}_{BS}$, the angle $\theta_{CS:BS}$ between $\mathbf{n}_{CS}$ and $\mathbf{n}_{BS}$, and the angles $\theta_{SW:CS}$ and $\theta_{SW:BS}$ between the solar wind flow direction and each of the two normals. 

The magnitude of the events was measured by two parameters -- the normalized field jump across the current sheet \citep{schwartz00},
\begin{equation}\label{eq6}
\Delta B_{12} = \frac{|B_1 - B_2|}{\mbox{max}(B_1, B_2)},
\end{equation}
and the maximum normalized amplitude of the field decrease in the core region,
\begin{equation}\label{eq7}
\Delta B_0    = \frac{B_{12max} - B_{0min}}{\mbox{max}(B_1, B_2)},
\end{equation}
in which $B_{12max}$ is the largest B-field magnitude observed in both pre- and post- sectors and $B_{0min}$ is the smallest field inside the core region.

To quantify the efficiency of the current sheet - bow shock interaction, we estimated the ratio between the transit velocity of the event, $V_{tr}$, and the local gyrovelocity of solar wind protons, $V_g$ \citep{schwartz83, schwartz00}:

\begin{equation}\label{eq4}
\left| \frac{V_{tr}}{V_g} \right|_{1, 2} = \frac{\cos (\theta_{SW:CS}) }{2  \cos (\theta_{SW:CS})  \sin (\theta_{B1,2:CS}) \sin (\theta_{CS:BS})}
\end{equation}
where indexes 1 and 2 apply to pre- and post sectors as usual. 

The parameters listed above were used to select candidate HFA events from the automatically detected set of magnetic field depression events satisfying the threshold condition (\ref{eq1a}). Final classification of events involved manual validation focused on the detailed shape of the magnetic field variation before, during and after the event, statistical properties of kinetic-scale magnetic field fluctuations indicative of an ion heating, and ion energy spectra.

\subsection{Fluctuation analysis}

Our analysis of MESSENGER's first flyby of Mercury has shown that the Hermean magnetosphere, as well as the surrounding region, are affected by non-MHD effects introduced by finite sizes of cyclotron orbits of the constituting ion species \citep{uritsky11}. Kinetic-scale magnetic fluctuations seems to play a significant role in Mercury's magnetosphere up to the largest resolvable timescale dictated by the signal nonstationarity. 

To investigate statistical properties of magnetic field fluctuations associated with the magnetic depression events, we used the method of higher order structure function (SF) generalized by \citet{uritsky11} for the case of strongly nonstationary signals. Using this tool, we compared magnetic turbulence inside and outside the detected events, and evaluated the ion crossover scale separating fluid-like and kinetic-like modes of behavior of solar wind plasma. 

The time-domain higher-order SF is defined as 
\begin{equation}\label{eq8} 
S_q(\tau)= \left\langle |\delta B_{\tau}|^q \right\rangle,
\end{equation}
in which $\delta  B_{\tau}$ are the differences of the studied turbulent field $B$ measured at time lag $\tau$, $\langle \cdot \rangle$ denotes averaging over all pairs of points separated by this lag, and $q$ is the order. The SF exponents $\zeta_q$ estimated from the scaling ansatz $S_q (\tau) \propto \tau^{\zeta_q}$ provide a detailed description of the turbulent regime under study. The second-order SF $S_2(\tau)$ plays a special role in statistical mechanics of turbulent media as a proxy to the band-integrated wavenumber Fourier spectrum \citep{biskamp03}. The power-law exponent $\beta$ of the spectrum is related to the SF exponent through $\zeta_2 = \beta -1$ under the assumption of linear space-time coupling. 

In order to study transient and/or spatially inhomogeneous solar wind fluctuations, we used local SF estimates made within a sliding window of width $W$. For each sliding window position, we computed a set of temporal SFs according to eq. \ref{eq8}, with $q = 2$ and $\tau < W/2$. The time-dependent shape of the resulting two-dimensional windowed structure function $S_2(\tau, t)$ was  represented as a continuous second-order time-period scalogram $\zeta_2(\tau, t)$: 

\begin{equation}\label{eq9} 
\zeta_2(\tau, t) = \frac{\partial \mbox{ log} \left[  \frac{1}{W - \tau +1 } \sum_{t'=t-W/2}^{t+W/2-\tau} {\left| \widetilde{B}(t')-\widetilde{B}(t'+\tau) \right|^2}   \right]   }{\partial \mbox{ log } \tau}.
\end{equation}
\noindent 

Here, $\widetilde{B}(t') = B(t') - \phi(t,t',\Delta)$ is the locally detrended magnetic signal, with $\phi$ being the quadratic polynomial fit to $B$ over the windowed time interval $t' \in [t-W/2, t+W/2]$, $\tau$ is the time scale, and $t$ is the running time variable reflecting the central position of the sliding window. We used quadratic detrending as the simplest way to compensate for the nonstationary trends reflecting spatial inhomogeneity of the traversed plasma structures. Following the approach proposed by \cite{matthaeus82}, the stationarity of the signal was verified based on the ergodic theorem for weakly stationary random processes \citep{monin75}. The partial derivative in the above equation is evaluated from the local least-square linear regression slope of the $S_2(\tau)$ dependence in the log - log coordinates for each sliding window. 

The continuous SF scalogram technique defined by eq. (\ref{eq9}) is essentially different from the widely used wavelet-based of Fourier transform-based dynamic spectrogram techniques (see e.g. \cite{alexandrova06, boardsen09}) as it allows analysis of temporal variations of the scaling structure of magnetic fluctuations rather than their spectral amplitudes. As shown in the next section, the SF scalogram provides evidence for drastically different turbulent plasma environments inside and outside the HFA events. In the absence of relevant particle data, this piece of information turns out to be particularly useful.

Using the Taylor frozen-in flow condition which is usually fulfilled in the solar wind (see e.g. \citet{matthaeus05} and references therein) and assuming that the upper spatial scale of ion-kinetic turbulent regime is controlled by finite Larmor radius effects \citep{schekochihin07}, the ion gyro radius $\rho_i$ and the ion temperature $T_i$ can be evaluated through \citep{uritsky11}

\begin{eqnarray}
   \rho_i & \approx & V_{SW} \tau_i / 2\pi, \label{eq10} \\
   T_i & \approx & m_i (V_{SW} \tau_i / \tau_{ci})^2, \label{eq11}
\end{eqnarray}
\noindent 
in which $\tau_i$ is the largest temporal scale at which the power-law slope of locally estimated SF is consistent with ion-kinetic $\zeta_2$ values, typically in the range 1.3 - 1.5 depending on the underlying dispersive wave mode (usually kinetic Alfv\'{e}n waves or whistler branches with secondary lower hybrid activity), and the turbulence type (i.e., a weak or strong) -- see, e.g.,
\citet{yordanova08, eastwood09, sahraoui09}. Compressional corrections further increase the kinetic-scale exponents \citep{alexandrova08}. For practical purposes, it is sufficient to use a simplified condition $\zeta_2 \approx 1$ to identify $\tau_i$ \citep{uritsky11}.

\section{Case studies} 

In this section, we present a subset of magnetic field depression events exhibiting magnetic and kinetic HFAs signatures, as compared to several non-HFA events associated with passages of helio current sheets (HCSs). All the events were initially detected automatically, after which the boundaries of their pre-, post- and core sectors were adjusted manually to better match their magnetic field profiles not captured by the default definitions. The parameters of the HFA and HCS events are summarizes in Tables \ref{table1} and \ref{table2}, correspondingly.

\subsection{HFA-like events}

Ten HFA-like events were identified during the studied 180 day period. Below we provide detailed portraits of each of these events.

\begin{figure*}
\noindent\includegraphics*[width=9 cm]{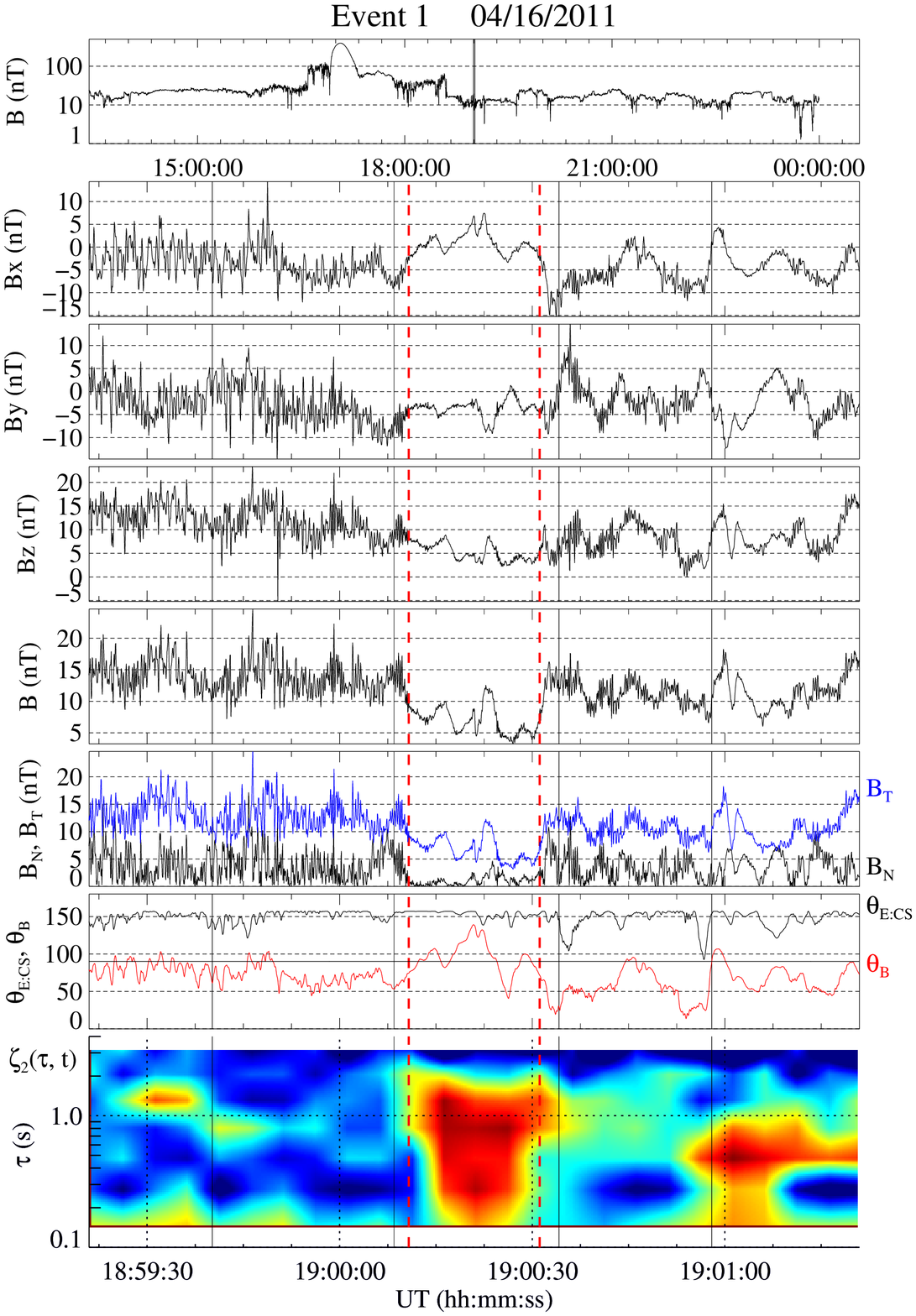}\noindent\includegraphics*[width=9 cm]{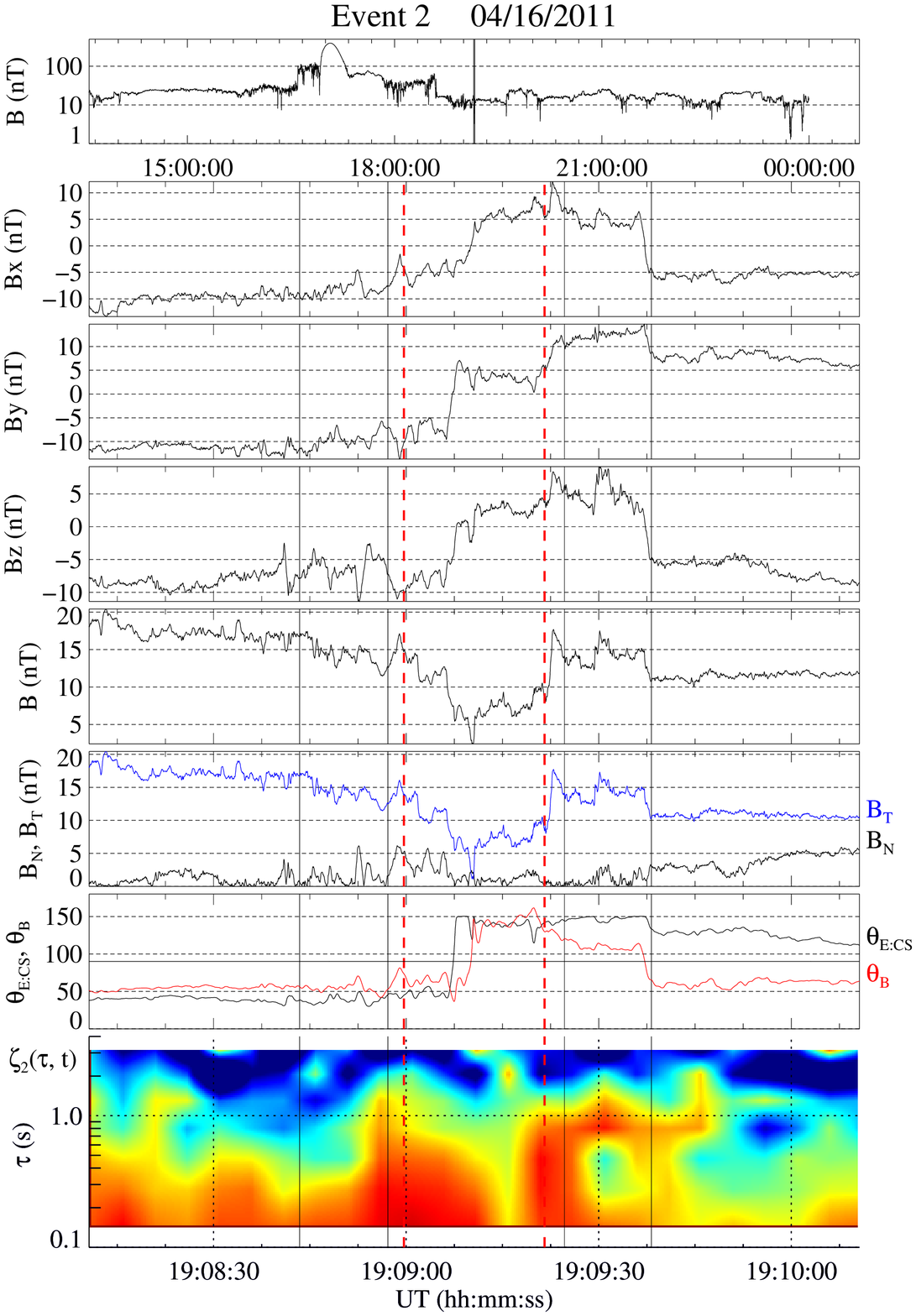}
\includegraphics*[width=4 cm]{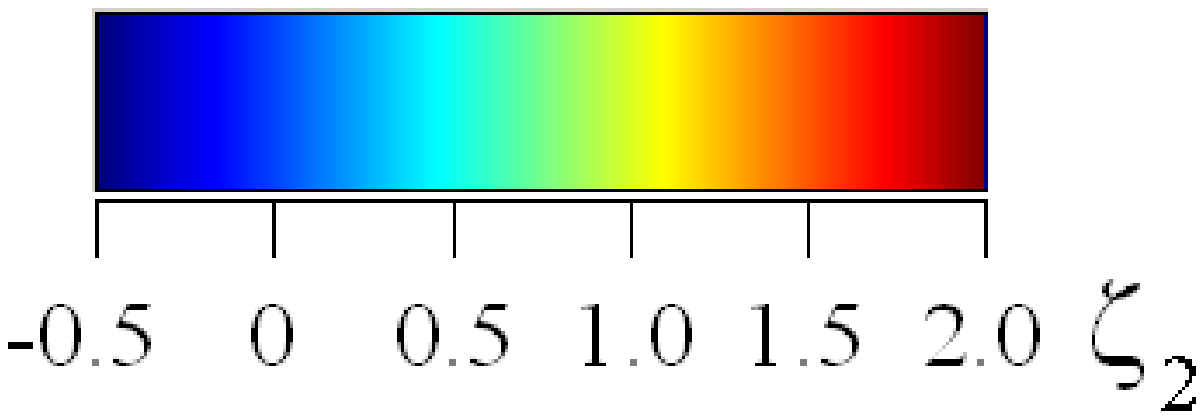}
\caption{\label{fig4} Magnetic and turbulent signatures of two hot flow anomaly events observed on April 16 2011. The top panels shows large-scale variations of the B-field magnitude $B$, with the black vertical line marking the timing of the event. The remaining panels (from top to bottom) show  zoomed-in plots of the $B_x$, $B_y$ and $B_z$ MSO magnetic field components; the B-field magnitude $B$; the tangential ($B_T$) and normal ($B_N$) magnetic field components in the current sheet coordinate system determined using eq. (\ref{eq3}); the angle $\theta_{E:CS}$ (black line) between the motional E-field and the current sheet normal and the cone angle $\theta_B$ (red line) between the magnetic field and the anti-sunward direction; the SF scalogram showing  temporal evolution of the second-order structure function exponent $\zeta_2$ estimated at different temporal scales $\tau$. The $\zeta_2$ color coding is the same for all the scalogram plots presented in this paper. Red dashed vertical lines show the boundaries of the core region, black solid lines mark the pre- and post sectors of the event.}
\end{figure*}

{\bfseries Event 1} (Fig. \ref{fig4}, left) occurred on April 16, 2011 just outside of the bow shock boundary, close to the noon-midnight meridian plane. The event was centered at about 19:00:22 UT and characterized by a normal motional electric field component pointed toward the current sheet in the post sector only ($\theta_{E2:CS} = 151^{\circ}$). Both leading and trailing edges of the event showed mild magnetic field enhancements, with the trailing edge lasting twice as long compared to the leading edge (respectively $\sim 3$ and $\sim 7$ s, or $\sim$ 1400 and 3200 km at a nominal solar wind speed of 450 km/s ). The more pronounced trailing edge may reflect a more the favorable E-field orientation on that side \citep{thomsen93}. The described features coexist with an irregular ULF oscillation which precludes their more accurate analysis. The event shows a small B-field rotation angle $\theta_{B1:B2}$ of $\sim 20^{\circ}$ and a quasi-parallel magnetic field alignment relative to the local bow shock normal. The core region of the event shows significantly reduced B-field magnitude ($\Delta B_0 \sim 1.1$) lasting for about 20 s, and a noticeably sharper trailing wall consistent with stronger plasma compression expected on the side exposed to the inwardly directed motional electric field. The normal magnetic field is close to zero and stays fairly constant inside the core region, and is 3-5 times lower than the tangential B-field both in the core and in the surrounding plasma environment, an indication of a nearly perpendicular shock \citep{paschmann88}. The jump in the tangential field $B_T$ by a factor of $\sim 2.5$ at the inner wall of the trailing edge implies a jump in the plasma density as required for such shocks. 

The core of event 1 shows a very clear and well-localized enhancement of kinetic-scale magnetic turbulence (bottom panel), with the ion kinetic crossover marked by the yellow color in the chosen color coding rising up to $\tau_i \sim 2$ s during the event. The kinetic crossover is not resolved by the MAG instrument in the ambient plasma region, suggesting that $\tau_i$ was increased by at least a factor of 10 in the core of this HFA. The observed change implies a proportional increase of the proton gyro radius which can not be explained by the much more modest drop of the average field magnitude during the event, unless it was accompanied by a plasma heating. By using eq. (\ref{eq11}) with $V_{SW}=450$ km/s, we estimate the proton temperature in the core region to be $\sim 95$ eV, or about $1.1 \times 10^6$ K, which is a factor of 5 greater than typical solar wind temperature. The obtained value is in agreement with the temperatures observed in the central region of terrestrial HFAs ($\sim 10^6 - 10^7$ K \citep{thomsen88, paschmann88, schwartz00}), and is three times higher than the typical proton temperature in the ambient solar wind plasma at Mercury's orbit \citep{baker09, uritsky11}. 

\begin{figure*}
\begin{center}
\noindent\includegraphics*[width=9 cm]{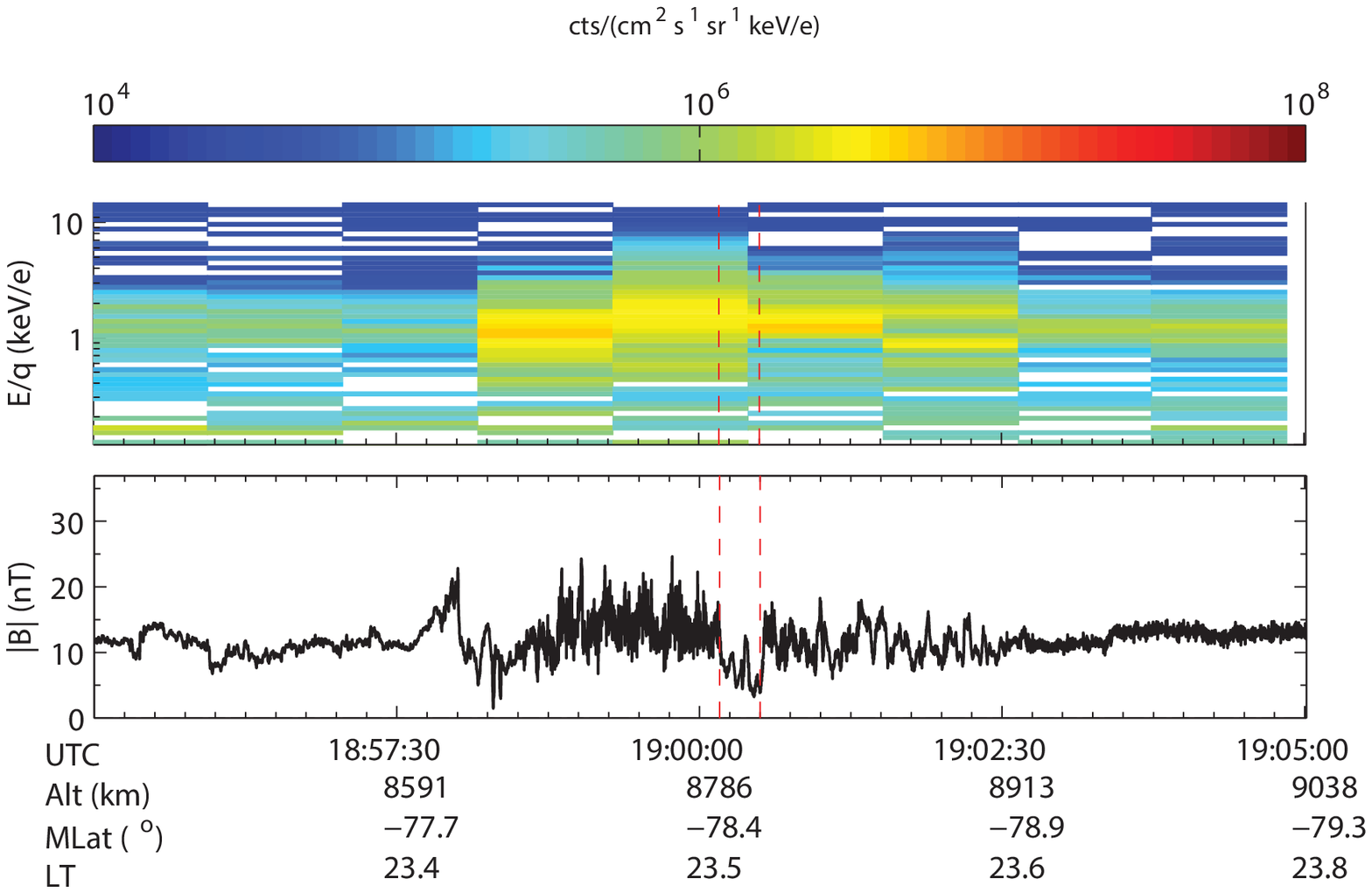}
\noindent\includegraphics*[width=9 cm]{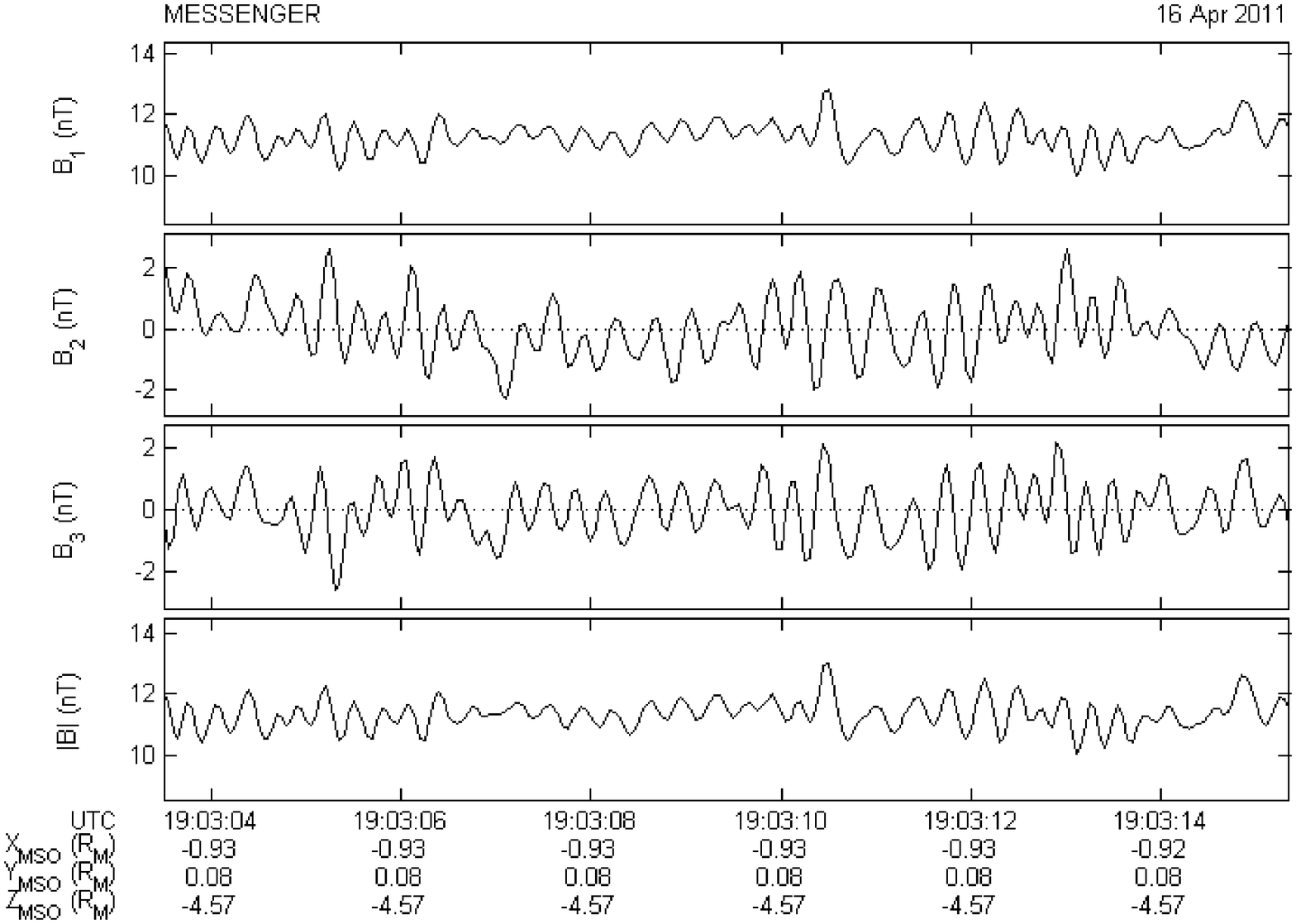}
\end{center}
\caption{\label{fig4_1} Top panel: Combined plots of FIPS and MAG observations of the hot flow anomaly event 1 shown in Fig. \ref{fig4}(left). The core region of the event is marked with red vertical lines. Bottom panel: left-hand polarized 3-4 Hz wave activity after the event. The right hand plot is given in field-aligned coordinates, where $B_1$ is the direction of the mean field over the plotted interval, and $B_2$ and $B_3$ are the transverse components, with the corresponding unit vectors $\mathbf{n_1} = [-0.85, 0.51, 0.13]$, $\mathbf{n_2} = [0.52, 0.85, 0.00]$, and $\mathbf{n_3} = [-0.11, 0.066, -0.99] $. }
\end{figure*}

Event 1 is supported by relevant particle observations. FIPS spectra show clear particle energization associated with this HFA. The FIPS spectra were averaged over the instruments angular field of view, 1.4 $\Omega$, of which 0.4 was obscured. During the presented observation interval, the FIPS $E/q$ range was set to 0.046 - 13.60 keV/q, and the scan time was 64 s. The color-coded plot in Fig. \ref{fig4_1} (top panel) shows the energy versus time spectrogram of the averaged differential proton flux during event 1. There is a noticeable increase in the ion energy both preceding and following the HFA, with the strongest energization roughly consistent with the position of the core region of the event. 
Although the low time resolution of FIPS data prevents a more detailed analysis of the structure of this event, it is sufficient for determining that a hotter ion population may be present at the HFA. The arrival directions measured by FIPS are consistent with the appearance of a beam-like flow, although the statistics are too limited to be analyzed quantitatively.

The angle between the motional electric field and the current sheet stayed almost constant in the pre-sector of event 1 (Fig. \ref{fig4}, left). It began to show transient departures from the original value of $\sim 150$  degrees in the post-sector of the event with the motional field being almost parallel to the current sheet plane on several brief occasions. These rotations had an irregular recurrence period of $\sim 7-15$s and could be an indication of unstable plasma conditions which in turn led to the strong ULF wave oscillations seen a few seconds after the event. These low frequency waves had a period of $\sim$ 10 seconds (0.1 Hz), wave amplitude of $\sim$ 5-10 nT, a low level of coherence, and a predominant left-hand polarization. 
A comparison of the minimum variance direction and the field-aligned direction suggests a wave normal angle of $\sim$ 10-15 deg. A second high frequency pulsation, around 3-4 Hz, was present simultaneously.  These waves were also left-hand circularly polarized (Fig. \ref{fig4_1}, bottom group of panel) and were observed from the edge of the core region until $\sim$ 19:07 UT. The wave intensity was varying on short time-scales and had a behavior similar to that of the electromagnetic ion-cyclotron waves. As the magnetic field in this region is nearly parallel with the bow shock normal, the wave excitation may be driven by back-streaming ions from the bow shock, along with escaping hot ions from the HFA, which also is consistent with the hot ion population observed over the wave interval.

{\bfseries Event 2} (Fig. \ref{fig4}, right) occurred about 9 minutes after event 1 and also showed a significant reduction of magnetic field strength in the core region ($\Delta B_0 \sim 0.8$). The fairly small normalized change in the B-field across the current sheet ($\Delta_{12} <0.1$) makes it difficult to interpret the discontinuity underlying this event as a clean tangential discontinuity (TD). The current sheet features a rather small $\theta_{CS:BS}$ angle of less then 10 degrees revealing its parallel alignment with the bow shock surface. Both parameters are the lowest for the presented collection of events, and are not typical for terrestrial HFAs. In addition, the event features a nearly perpendicular geometry ($\theta_{B1,2:BS} \approx 90^{\circ}$) with no strong magnetic connection with the bow shock. In spite of the unusual orientation, event 2 has a consistent ``toward'' orientation of the motional electric field both in the pre- and post-sectors resulting in well-defined and almost symmetric trailing and leading edges revealing compressed plasma regions around the anomaly. On the other hand, the E-field direction outside of the core region remains effectively unchanged. The observed $\theta_{E:CS}$ dynamics could be consistent with an interplanetary discontinuity surrounded by relatively quiet plasma regions, and is not characteristic of strong HFAs which tend to be embedded in a more violent foreshock environment. The SF scalogram shows a mild increase of the ion-kinetic time scale at the leading edge of the event (from 0.3 to $\sim 2$ s), with the estimated core region temperature of about 30 eV suggesting only a marginal heating. This is consistent with the rather large ratio $\left| V_{tr} / V_g \right| \sim 3$ at both sides of the event. Based on this estimate, the event can be classified as a moderately active HFA-like current sheet which moved too fast to experience any significant gyrokinetic heating.

\begin{figure*}
\noindent\includegraphics*[width=9 cm]{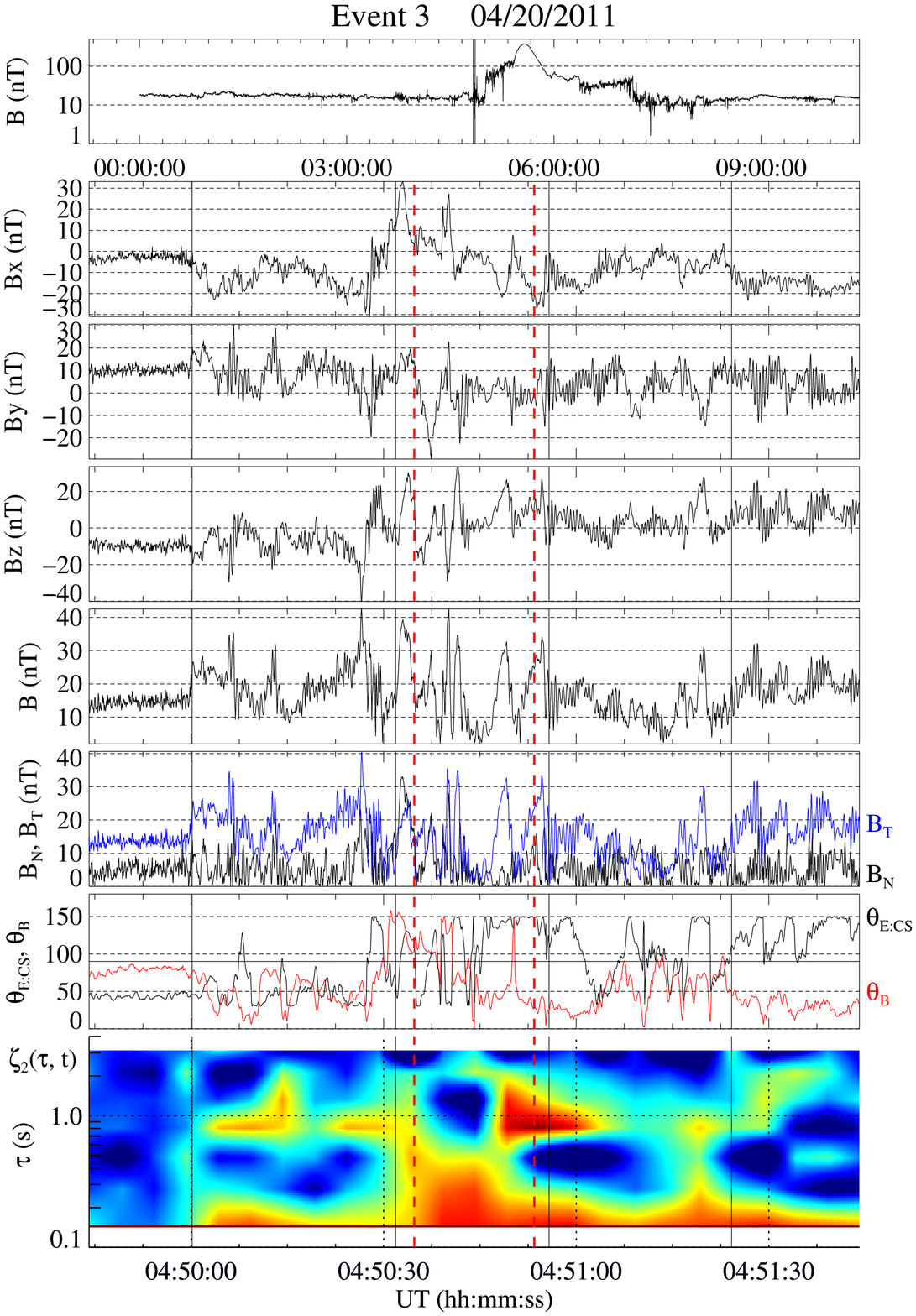}\noindent\includegraphics*[width=9 cm]{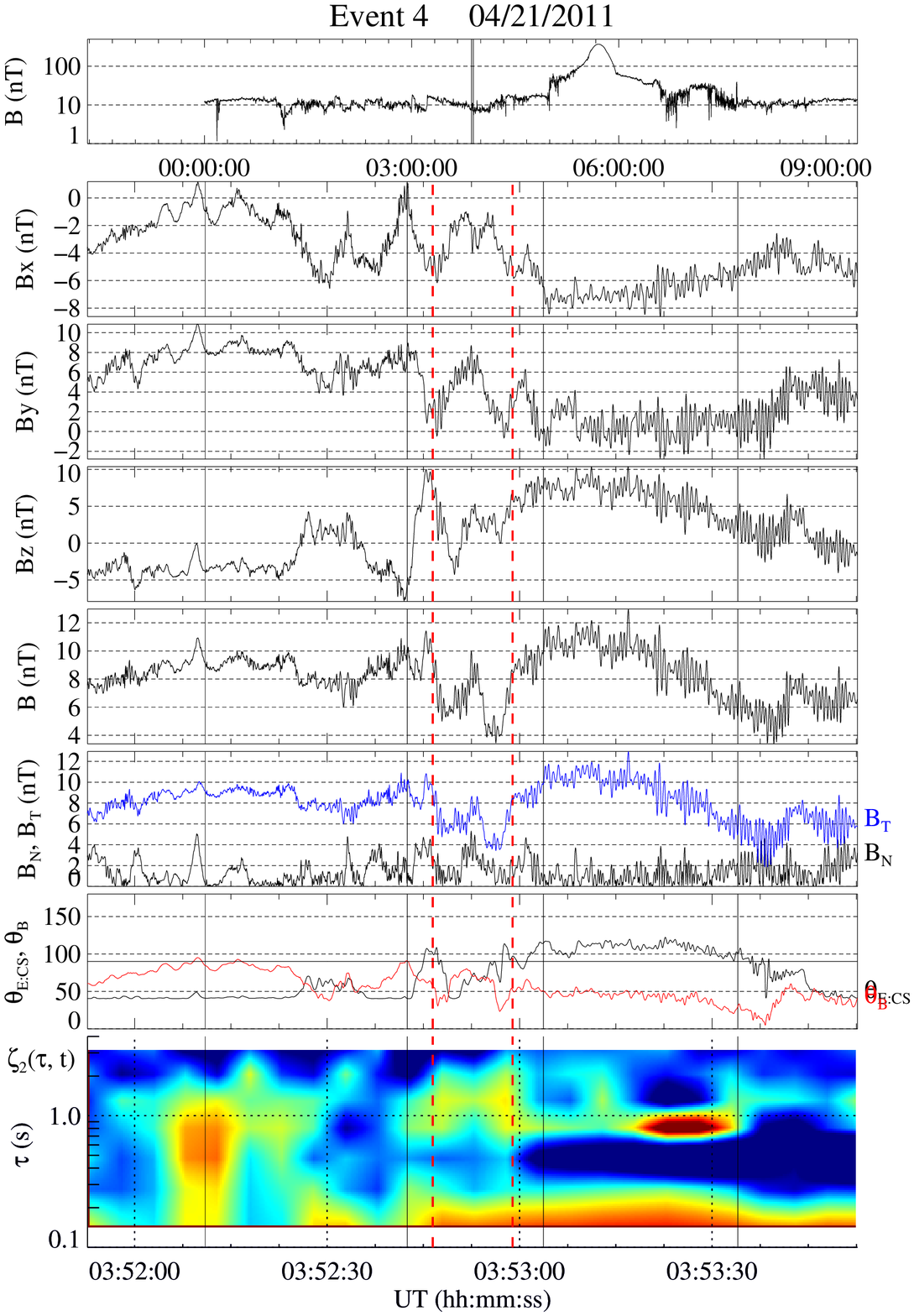}
\caption{\label{fig5} Hot flow anomalies observed on April 20 and April 21, 2011.}
\end{figure*}

{\bfseries Event 3} (Fig. \ref{fig5}, left) was observed at 04:50:30-04:51:00 UT on April 20, 2011. It features more pronounced HFA signatures compared to the previous event. Event 3, as well the rest of the events discussed below in this subsection, had properly oriented motional electric fields pointing toward the current sheet on either sides of the event. The event shows a fairly large difference in the B-field magnitude in the pre- and post-sectors described by $\Delta B_{12} \sim 0.4$ making it an almost certain TD \citep{neugebauer84}. The magnetic field undergoes several rapid directional changes inside the core region revealing multiple embedded current sheets. A significant normal magnetic field component associated with these changes signals an arbitrary  orientation of the small-scale current sheets relative to the main current sheet. The finest temporal scale of the embedded discontinuities is of the order of a tenth of a second which translates into the spatial scale of $\sim$ 40-50 km for typical solar wind conditions. These structures have a kinetic origin and could be supported by several distinct ion populations, deflected at different angles by the bow shock boundary. The intermittent structure of the discussed event resembles short large-amplitude magnetic structures (SLAMS) which are commonly observed at a quasi-parallel terrestrial bow shock \citep{schwartz92}, and recently in the foreshock of Venus \citep{collinson12a}. The structure of event 3 is similar to SLAMS embedded within a boundary with regions of considerable heating and deceleration. The time scale, polarization parameters, and other characteristics of SLAMS observed at Earth a suggestive of their growth out of ULF wave packets \citep{schwartz92} which are commonly found in HFA cavities. The wave field of less evolved HFAs can be quite complex \citep{tjulin08}. Event 3 is likely to belong to this category because of its location near the subsolar point suggesting a recent initial interaction with Hermean bow shock. All other HFA events considered in this section were located further downstream and therefore had more time to develop, which may explain their less turbulent core region environment. 

The core region of event 3 has well-developed edges with sharp field gradients indicative of strong plasma compression. In the wake of the event, there is a coherent oscillatory activity with a frequency of about 2 Hz which is $\sim 10$ times higher than the local proton gyro frequency. The mechanism of this post-sector wave activity which we saw in several other Hermean HFAs remains to be understood. The SF scalogram shows a dramatic increase of the ion crossover scale in the core region of the event, up to $\sim 2$ s near the inner side of the trailing edge. When plugged into eq.(\ref{eq11}), this fluctuation scale predicts a proton temperature of $\sim 470 $ eV, or about $5.5 \times 10^6$ K, which is comparable with typical HFAs seen at Earth. The scalogram also suggests that the heated, and presumably strongly deflected core particle population ``leaks'' through the trailing edge into the portion of the post-sector adjacent to the event. If this leakage does take place, it can play an important part in the excitation of a plasma instability underlying the 2 Hz wave oscillation in this sector. 

\begin{figure*}
\begin{center}
\noindent\includegraphics*[width=9 cm]{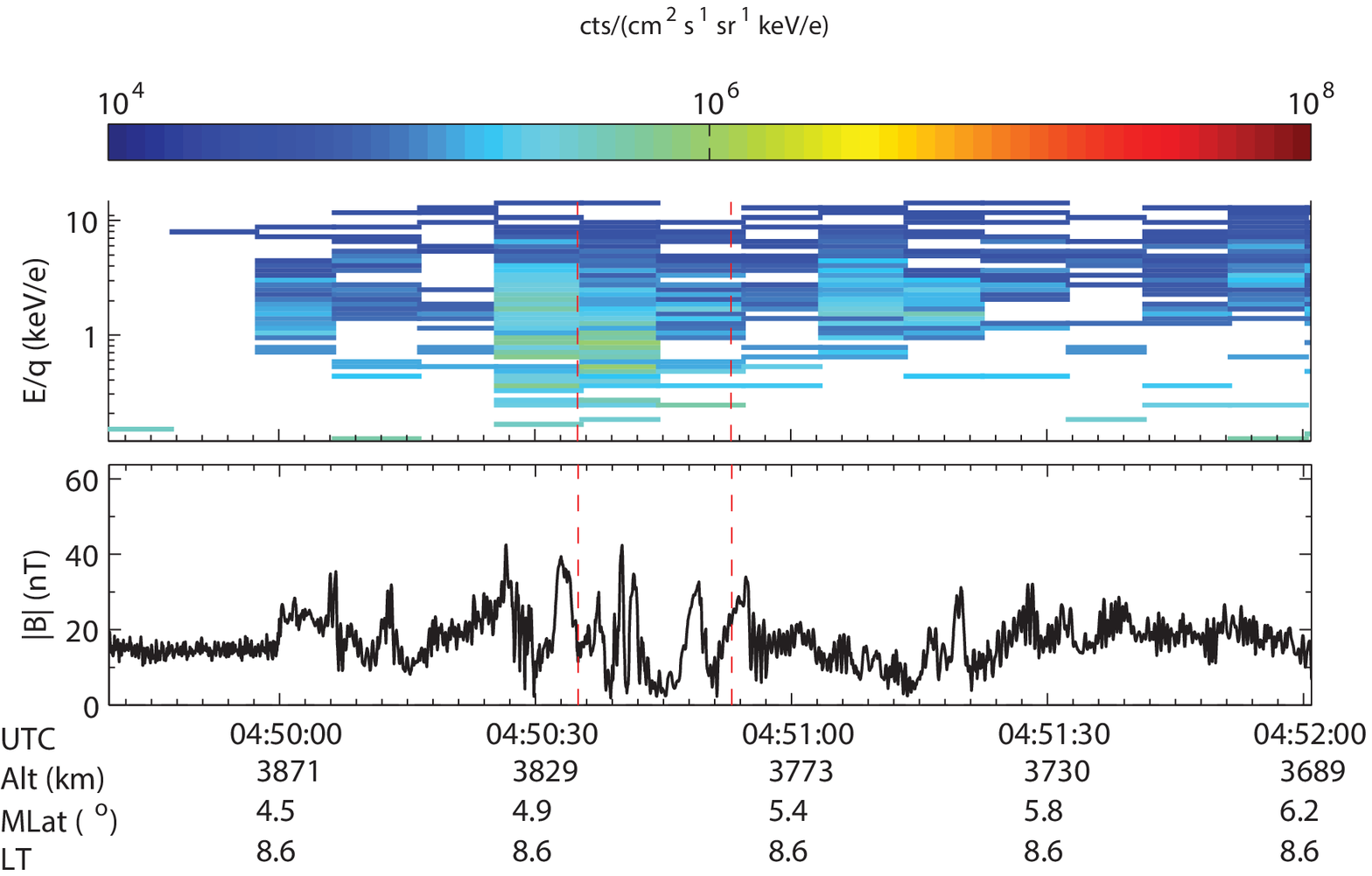}
\end{center}
\caption{\label{fig5_1} Combined plots of FIPS and MAG observations of the hot flow anomaly event 3 shown in Fig. \ref{fig5}(left). The core region of the event is marked with red vertical lines. }
\end{figure*}

During event 3, FIPS operated in its fast scanning mode ensuring a 8-second time resolution. At this sampling time, it is possible to match the heated plasma region with the magnetic signature of the HFA (Fig. \ref{fig5_1}). In can be seen that the event is associated with a broadened range of energies ($\sim$ 100 eV - 10 keV) taking place in the core region and at the trailing current sheet. The presented FIPS spectrogram indicates a substantial plasma heating. 

{\bfseries Event 4} (Fig. \ref{fig5}, right) occurred at about 03:52:45 - 03:53:00 UT on April 21, 2011. Despite the symmetric E-field geometry providing favorable HFA conditions on either side of the cavity, only the leading edge of this event has a noticeable compression which lasted for $\sim 3$ s (or approximately 1350 km in size for $V_{SW} = 450$ km/s). The magnetic field depression in the core regions is characterized by a relatively small normalized magnitude $\Delta B_0 \sim 0.5$. The event shows a moderate enhancement of kinetic-scale turbulence which nevertheless caused a sizable jump in $\tau_i$ from $\sim 0.2$ s in the late pre-sector to more than 1 s in the core region. The predicted pre-sector temperature is an order of magnitude lower than expected for an ambient solar wind and may indicate a presence of strongly decelerated ions in front of the propagating event. The remarkably stable electric field angle $\theta_{E:CS}$ observed prior to the arrival of the event agrees with this interpretation pointing at a thermodynamically stable and unperturbed plasma medium. 

In the middle of the core region of event 4, there is an isolated pillar of enhanced magnetic field magnitude which is considerably larger that the spikes observed in the core region of event 3. Singular embedded structures such as the one identified inside event 4 are indicative of a partial flow recovery and are often observed in terrestrial HFAs \citep{schwartz95}. More recently, they have been found at Venus \citep{slavin09}.

\begin{figure*}
\noindent\includegraphics*[width=9 cm]{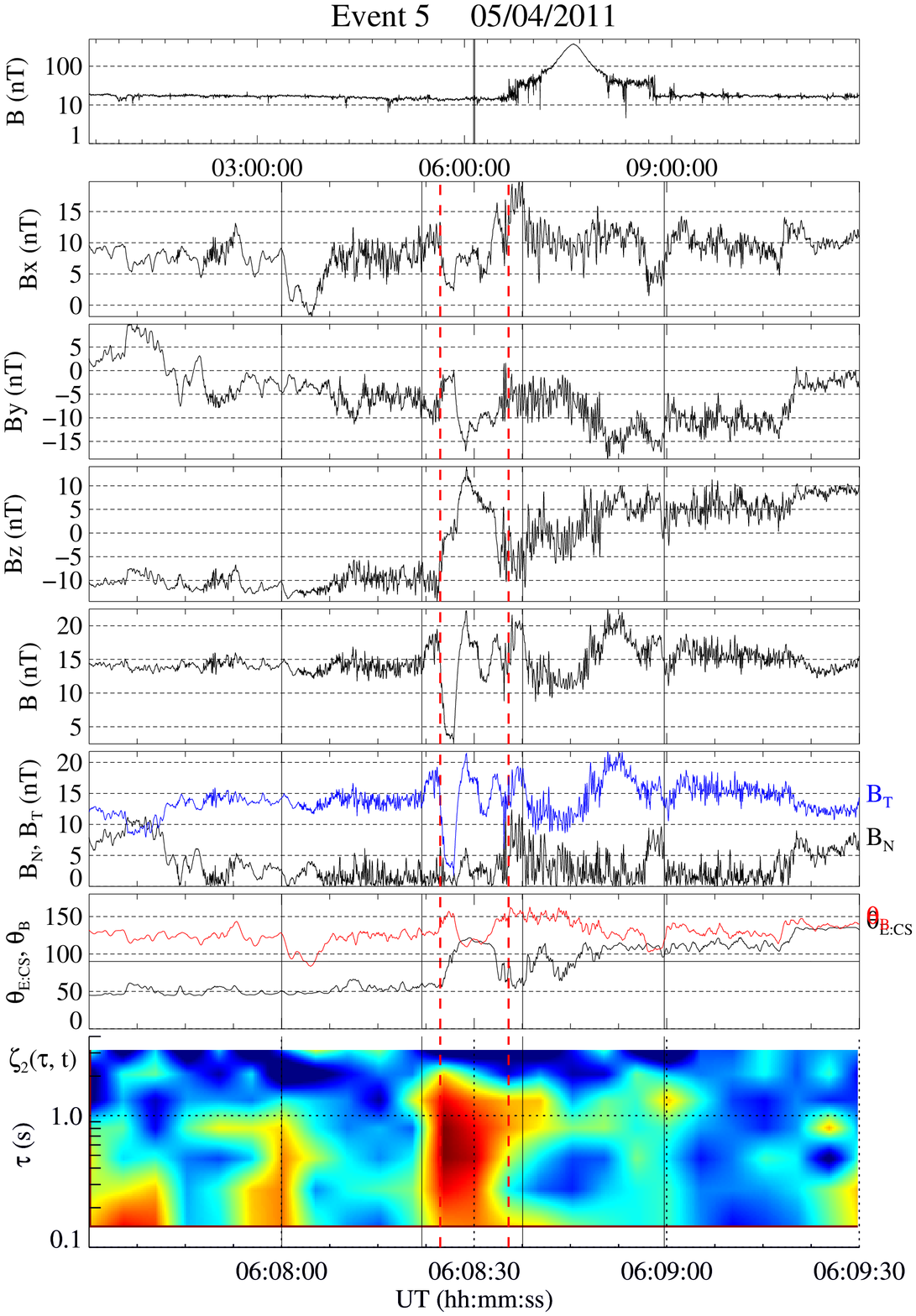}\noindent\includegraphics*[width=9 cm]{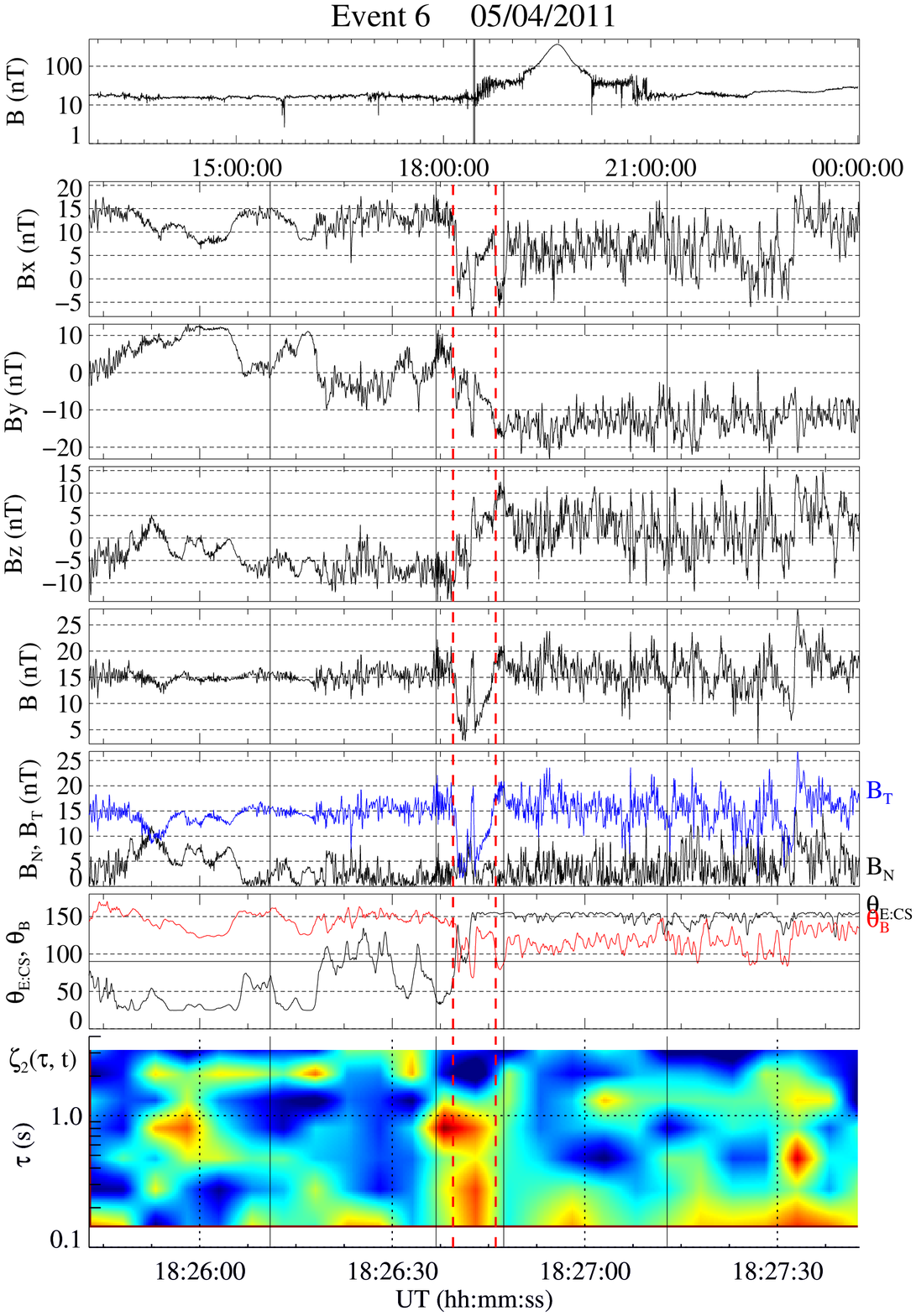}
\caption{\label{fig6} Hot flow anomalies observed on May 04, 2011.}
\end{figure*}

Events 5 and 6 (Fig. \ref{fig6}) were both observed on May 04 2012 during two subsequent MESSENGER orbits. The events were centered at 06:08:30 UT and 18:26:40 UT, correspondingly. In spite of the 12-hour time separation, the events have much in common. Their location is quite close to the equatorial plane in the dawn foreshock region with a quasi-parallel magnetic field geometry ($\theta_{B:BS} \sim 20-50^{\circ}$), and both events show classical compression signatures at the leading and trailing edges consistent with a convection E-field pointed inward on either side of the current sheets. The compressed edges are clearly seen in event 5 but are also identifiable in event 6.  Both events show pronounced transient heating signatures in the SF scalogram coinciding with the core regions of the events, and a similar magnetic field rotation angle of $60-70$ degrees. 

{\bfseries Event 6} (Fig. \ref{fig6}, right) exhibits a strong transient enhancement of the B-field magnitude in the middle of the core region analogous to that of event 4 (Fig. \ref{fig5}, right). The peak of this brief enhancement lasts for $\sim 0.7$ s, suggesting a spatial scale of a few hundred kilometers. Based on the $\tau_i$ estimate from the SF scalogram, the temperature of the plasma environment surrounding this structure is of the order of 50 eV, with a proton gyro radius of $\sim 140$ km. The discussed set of properties, especially the sharp inner walls of the core region, should be of kinetic origin. Interestingly enough, event 5 (Fig. \ref{fig6}, left) shows an even more significant variation of the magnetic field magnitude in the core region which, however, has a longer temporal scale ($\sim 2$ s) and a larger estimated size ($\sim 900 $ km ) compared to event 6. According to the SF scalogram, the core region temperature inside event 5 rises up to 360 eV, but due to the stronger average B-field, the predicted proton gyro scale is roughly the same as during event 6. Overall, events 5 and 6 are in a qualitative agreement with the observations of HFAs at Earth's quasi-parallel bow shock. 

Event 6 has a quite short core region duration ($T = 6$ s) although its travel- to gyro velocity ratio is fairly low ($\sim 0.42$ in the pre-sector). The low bow shock sliding velocity could explain the efficient heating of this compact event.

\begin{figure*}
\noindent\includegraphics*[width=9 cm]{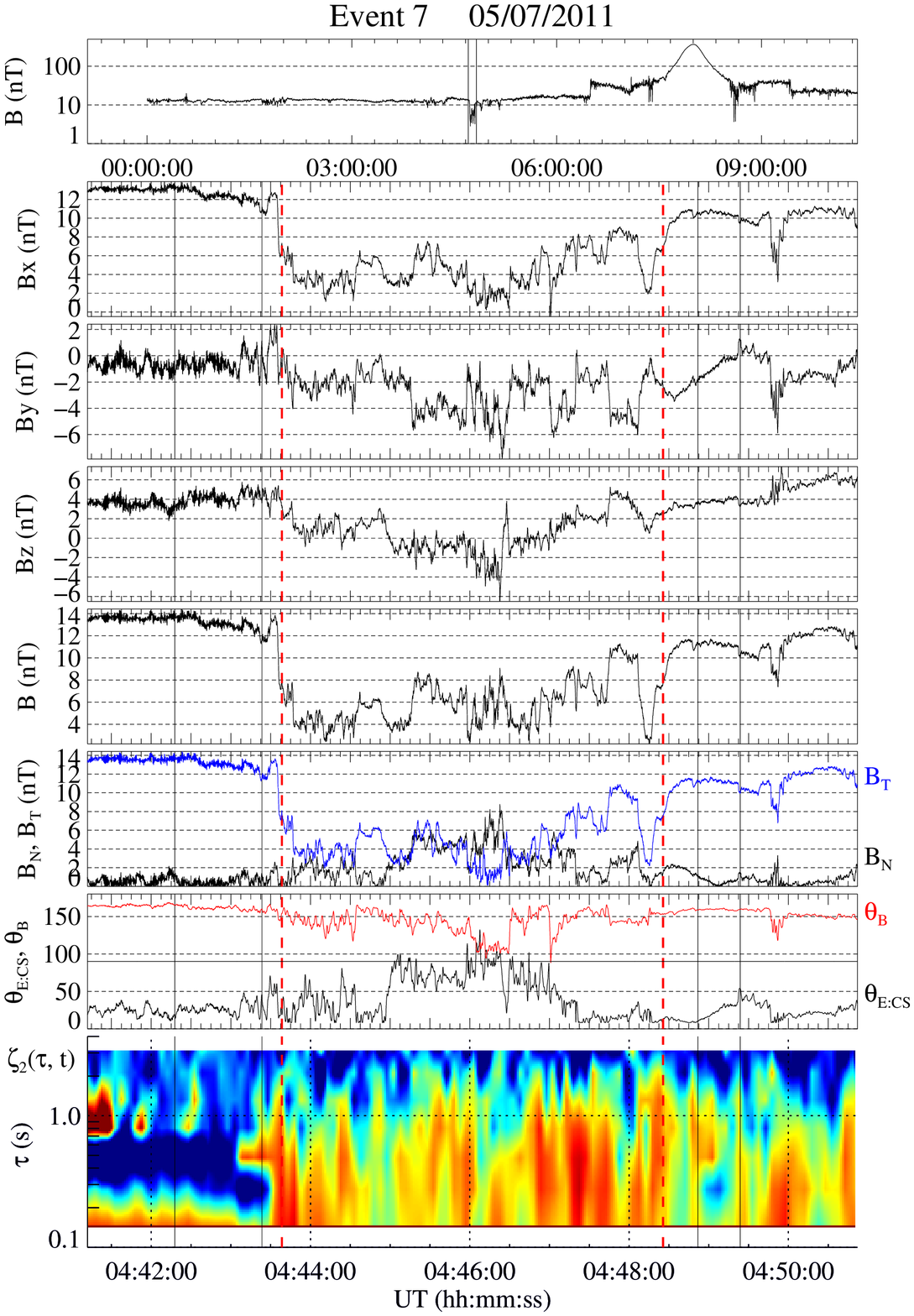}\noindent\includegraphics*[width=9 cm]{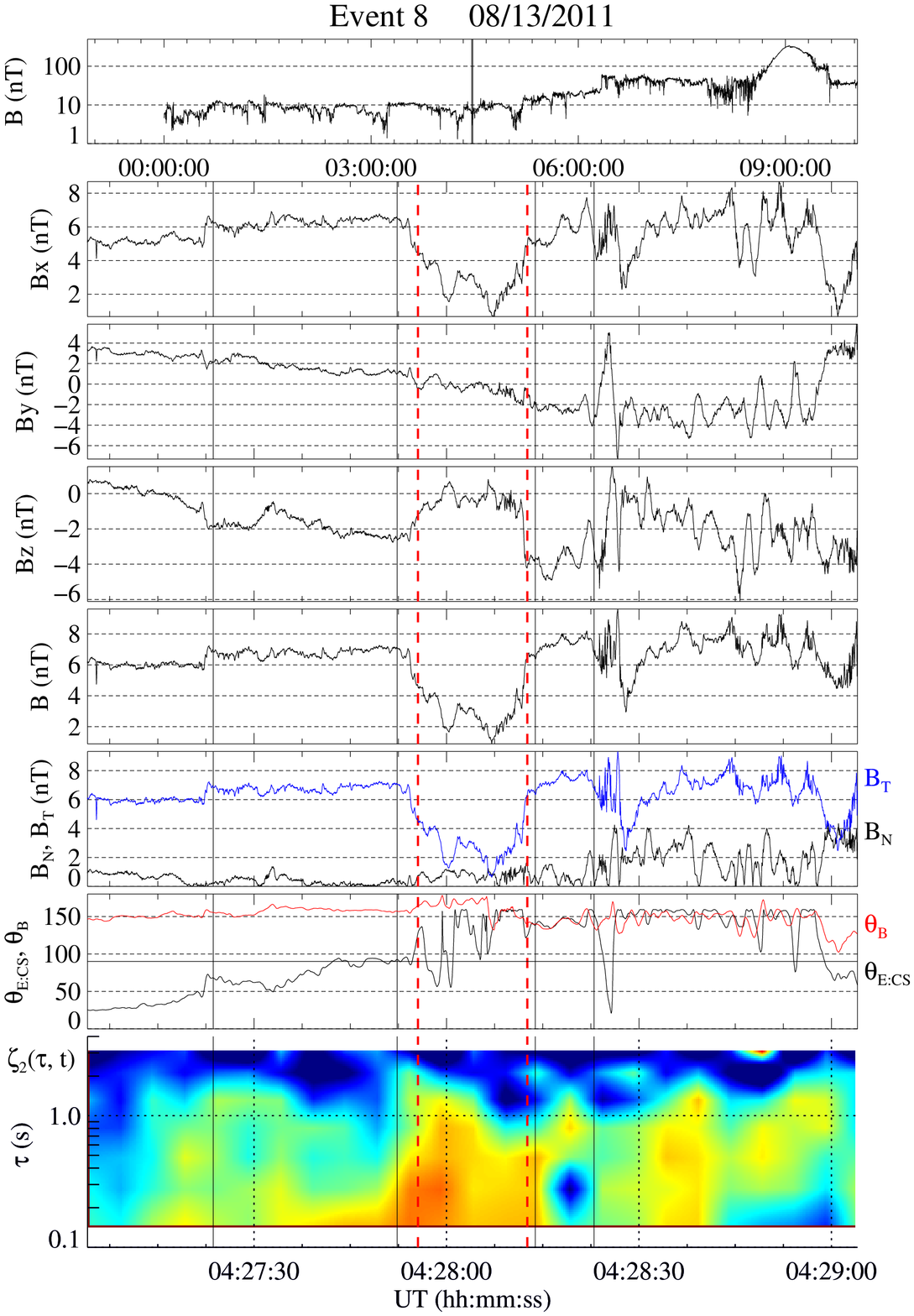}
\caption{\label{fig7} Hot flow anomaly - like events observed on May 07 and August 13, 2011.}
\end{figure*}

{\bfseries Event 7} was observed on May 07 2011 between about 04:43:40 - 04:48:30 (Fig. \ref{fig7}, left). It has the longest 
duration among the detected HFA events, with the core region passage lasting for almost 5 min. The ``toward'' motional electric field is observed in the pre-sector which also features a small but distinct compression edge. The plasma content of the core region is significantly nonuniform. According to the fluctuation scalogram, the leading edge is rather sharp and is associated with a transition from a high-frequency spiky noise (possibly of electrostatic origin \citep{singh07}) outside the event to a cross-scale turbulent cascade observed inside. At the trailing edge, the scaling structure of magnetic fluctuations changes in reverse order, but the time of this transition is not well defined. Irregular variations of the fluid-kinetic crossover seen inside the core region confirm the presence of multiple hot plasma layers separated by a substantially cooler plasma. The hottest ion population is observed near the trailing edge where $\tau_i$ reaches 1.0 - 2.0 s, which using eq. \ref{eq11} predicts a temperature range of $4-11 \times 10^5$ K, or 30 - 95 eV. 

Some of the magnetic field depressions inside the core region of event 7 are associated with considerable changes in the magnetic field orientation signaling small-scale current sheets. One of such embedded structures was encountered at 04:46:30. It was accompanied by an abrupt 50-degree change of the $\theta_B$ angle. Another rotation for an even larger $\sim$ 75 degrees angle was detected at about 04:47:00. These compact current-carrying  structures can be sites of separate heating events. Some of the local minima in the B-field magnitude are matched by substantial decreases of the normal component hinting at tangential discontinuities. The mutual arrangement of the embedded current sheets does not show stable periodicity and is likely to be shaped by MHD turbulence. 

Event 7 is characterized by a weak connection to the bow shock both before and after the core region, with $\theta_{B1:BS} \approx \theta_{B2:BS} \sim 70$ degrees, and is located relatively far from from the nominal bow shock boundary ($d_{BS} \sim 1.6 R_M$). The  magnetic field rotation between the pre- and post-sectors is unremarkable. Assuming that the hot core region is passively advected with the nominal solar wind speed and taking into account the angle $\theta_{SW:CS}$ between the solar wind flow and the current sheet, the thickness of this HFA is about 7 $R_M$. The velocity ratio $| V_{tr}/V_g |$  is estimated to be $\sim 0.35$ before and after the event, implying that the transit speed was sufficiently slow for the ions to be transported along the shock. The long interaction time could help the development of this event in the quasi-perpendicular bow shock geometry untypical for HFAs. 

\begin{figure*}
\begin{center}
\noindent\includegraphics*[width=9 cm]{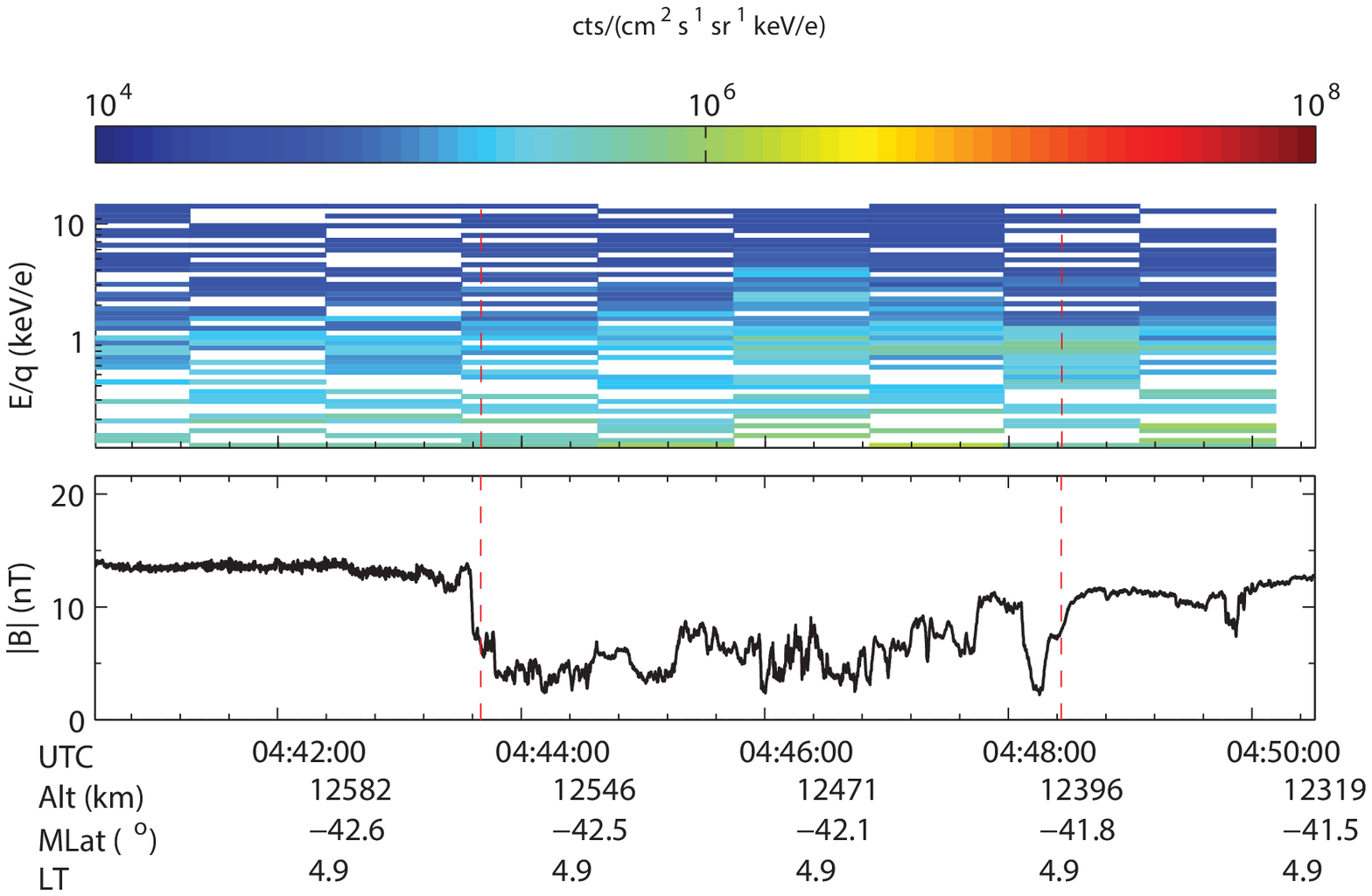}
\end{center}
\caption{\label{fig7_1a} Combined plots of FIPS and MAG observations of the hot flow anomaly event 7 shown in Fig. \ref{fig7} (left column). The core region of the event is marked with red vertical lines. }
\end{figure*}

Ion energization during event 7 is confirmed by FIPS measurements (Fig. \ref{fig7_1a}). The most energetic ion population was encountered close to the center of the core region, between 04:46:10 and 04:47:25. The heated plasma extends beyond the trailing event edge into the post-sector, in agreement with the washed-out trailing edge position on the SF scalogram as discussed above. 

{\bfseries Event 8}  was observed on August 13, 2011 (Fig. \ref{fig7}, right), with the core region center at $\sim$ 04:28:00. It features favorable orientation of the motional E-field at both sides of the current sheet. The post-sector shows a somewhat better magnetic connection with the bow shock compared to the pre-sector. The SF scalogram demonstrates a clear-cut fluctuation signature of ion temperature enhancement similar to that seen in other Hermean HFAs. The core region temperature increase is likely to be small, of the order of 20-30 eV, but it is statistically significant compared to the the ambient plasma temperature. The compression shoulders are apparently missing. Based on these signatures, event 8 could be classified as a proto-HFA event rather than a fully developed HFA, analogous to event 2 shown in Fig. \ref{fig4} (right). 

\begin{figure*}
\begin{center}
\noindent\includegraphics*[width=9 cm]{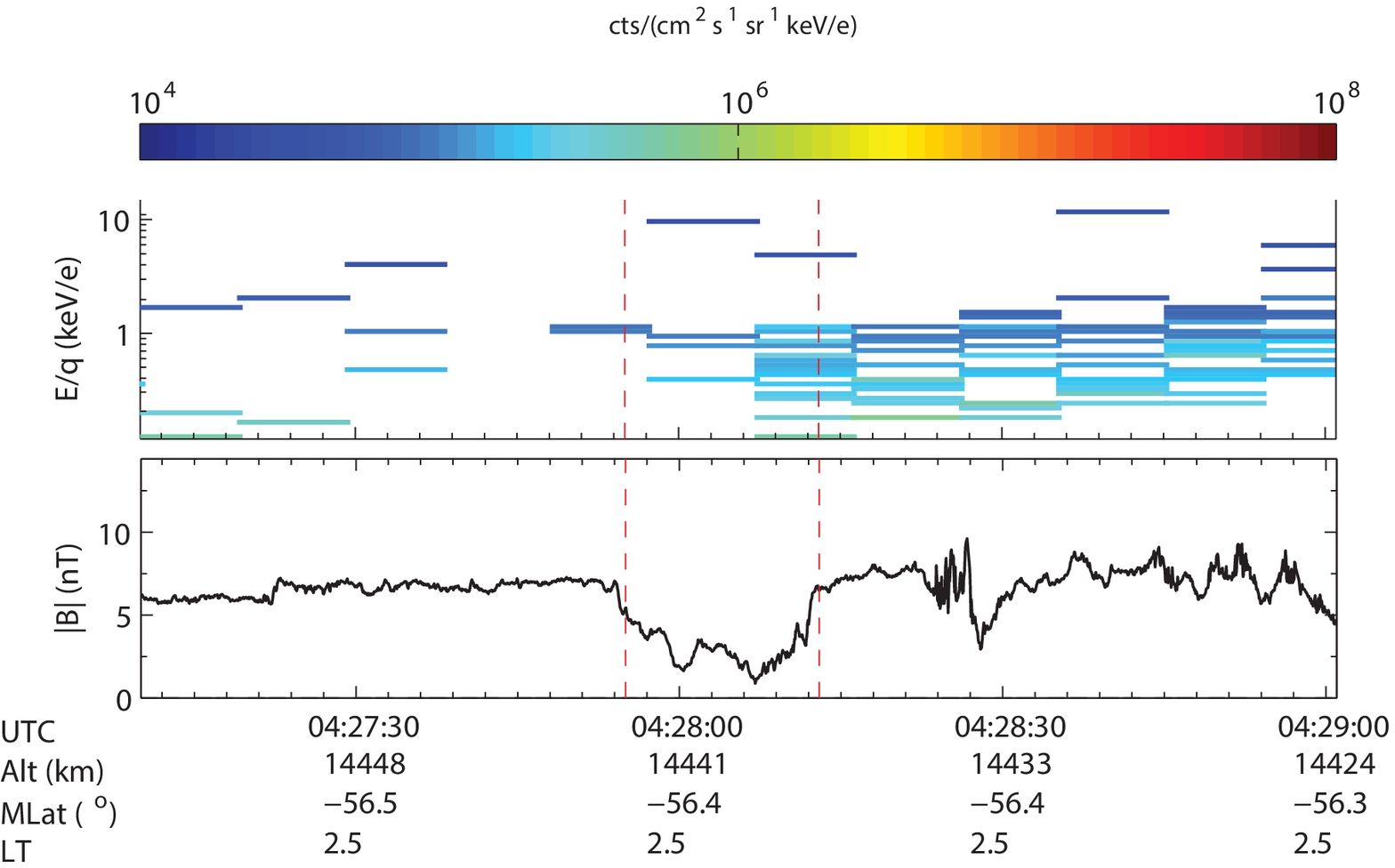}
\end{center}
\caption{\label{fig7_1b} Combined plots of FIPS and MAG observations of the hot flow anomaly event 8 shown in Fig. \ref{fig7} (right column). The core region of the event is marked with red vertical lines. }
\end{figure*}

The FIPS energy spectrogram of event 8 (Fig. \ref{fig7_1b}) is much like that of event 3. The spectrogram was obtained in the fast scanning mode and demonstrates a wide range of ion energies ranging from 100 eV to about 2 keV. This behavior is atypical for the solar wind and is consistent with the presence of a local plasma heating mechanism. The hotter post-sector of event 8 is in an agreement with the asymmetric shape of the turbulence scalogram predicting a larger ion kinetic scale in that sector compared to the pre-sector. 

\begin{figure*}
\noindent\includegraphics*[width=9 cm]{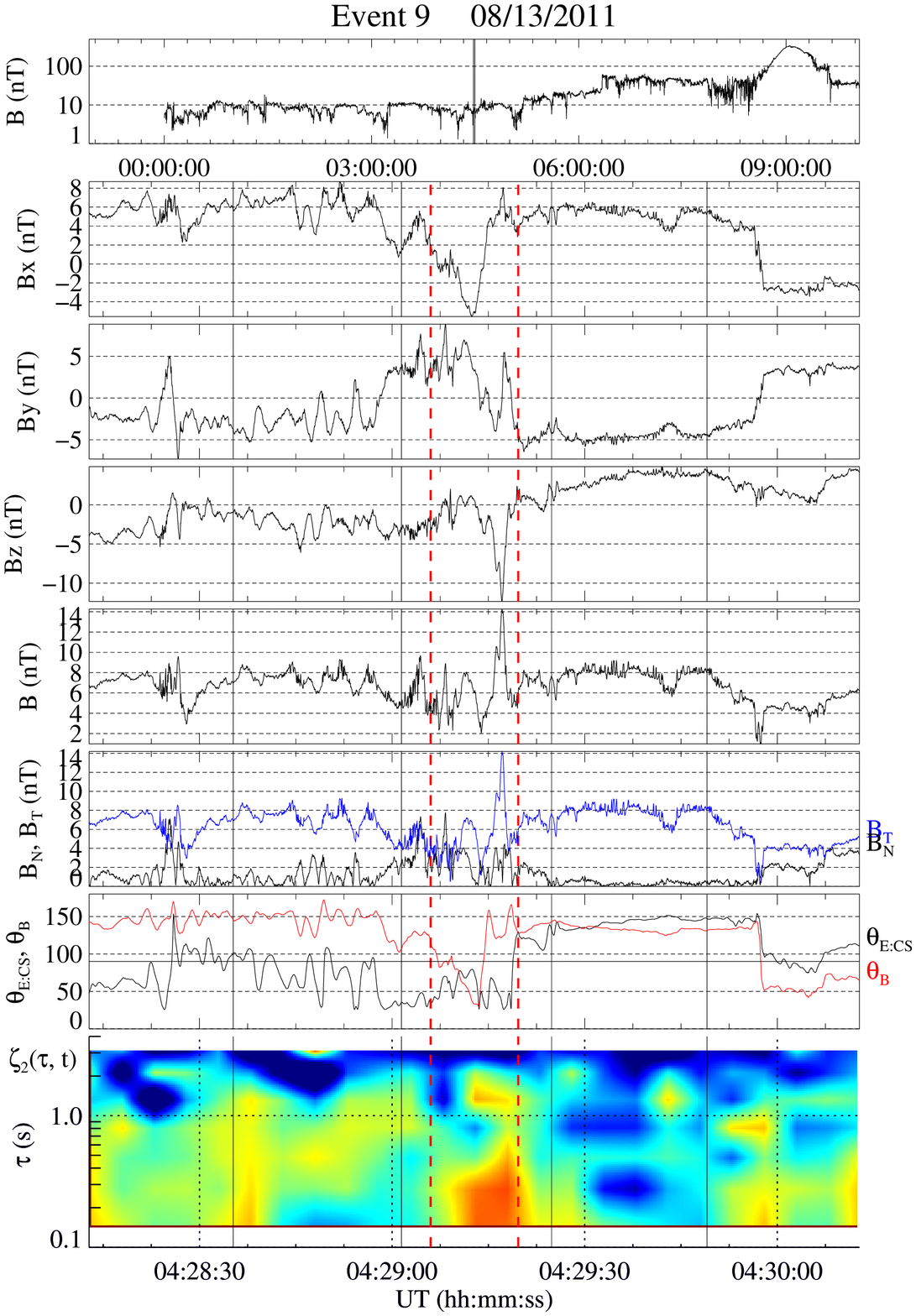}\noindent\includegraphics*[width=9 cm]{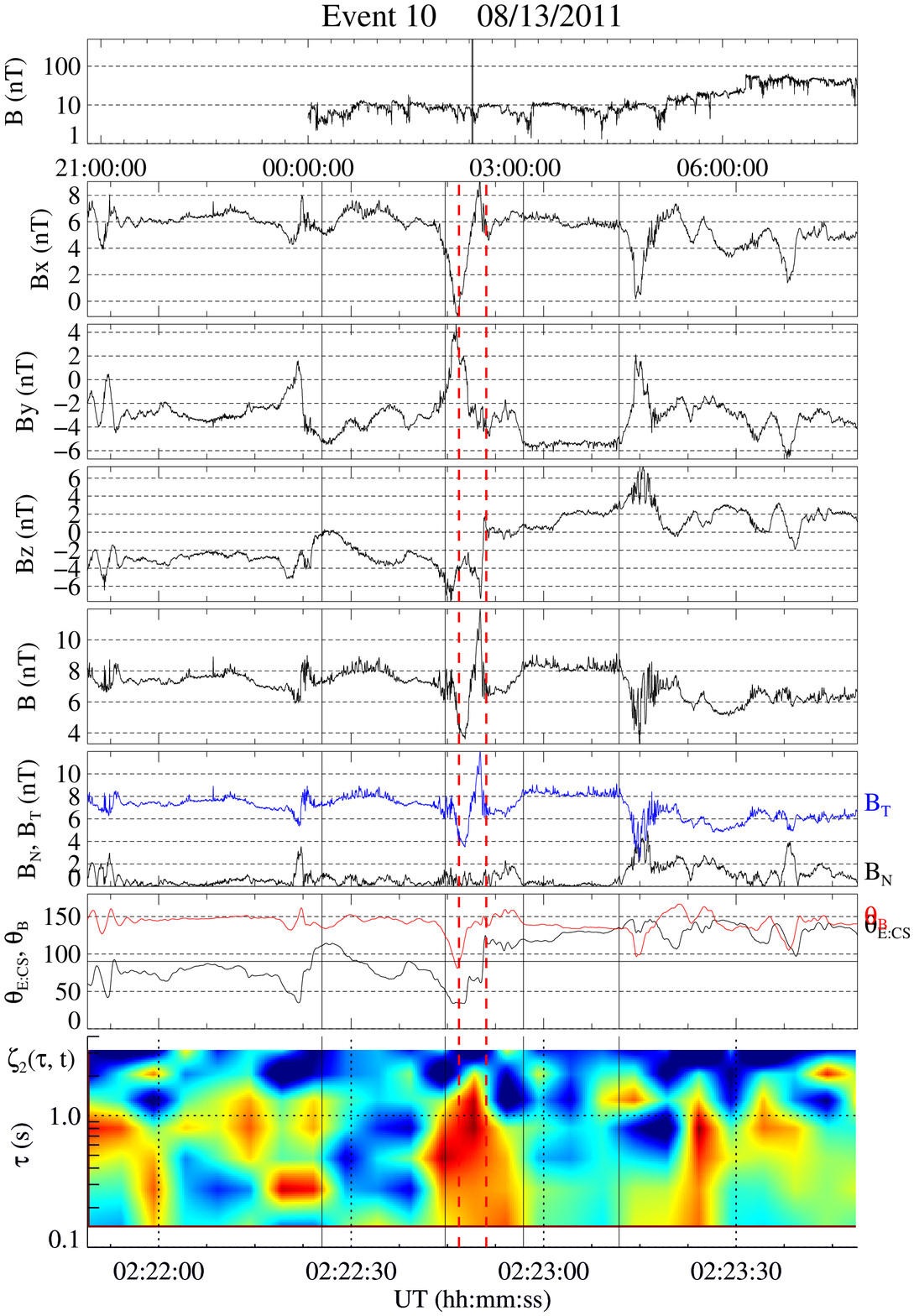}
\caption{\label{fig8} Two hot flow anomalies observed on Augusts 13, 2011.}
\end{figure*}

{\bfseries Event 9} (Fig. \ref{fig8}, left) shows a ``toward'' E-field orientation on both sides of the core region and has approximately that same duration as event 8. Event 9 has a consistent non-zero $B_N$ component in the core region, comparable with the tangential magnetic field, and demonstrates several B-field reductions inside the core region reminiscent of the embedded current sheet structures in event 7. However, event 9 involves a significant net magnetic field rotation of about 50 degrees. It shows no identifiable compression regions, although its trailing edge is rather sharp implying a thin current-carrying plasma layer. The post-sector of event 9 has a considerably weaker magnetic connection with the bow shock compared to the pre-sector, which leaves open the question of whether the trailing current sheet is due to an HFA-driven plasma expansion or is a pre-existing solar wind structure. The event was accompanied by a modest B-field depression ($\Delta B_0 \sim 0.37$) but a clearly-manifested increase of the ion-kinetic crossover time from $\tau_i \leq 0.6$ s at the inner edge of the core region to about $\tau_i \sim 1.3$ at its center, implying a 10-fold temperature growth. 

{\bfseries Event 10} took place at about 02:22:45 of Aug 13, 2011 (Fig. \ref{fig8}, right). It shows the shortest duration ($\Delta_t \sim 4$ s) over the entire set of HFA-like current sheets reported here, and is also characterized by the smallest relative reduction of the B-field magnitude. A clear ion energization footprint can be seen on the scalogram plot, with the ion-kinetic scaling extending up to $\tau_i \sim 2$ s inside the core region. The ion temperature predicted  for the crossover time and the observed field magnitude is $1.1 \times 10^6$ K, or 90-100 eV. A relatively small ($\sim 32$ degrees) but consistent magnetic field rotation observed outside the core region suggests that event 10 is embedded into a current sheet. This current sheet is nearly perpendicular to the nominal bow shock surface ($\theta_{CS:BS} \sim 100$ degrees) and has a stable magnetic connection with the bow shock in the pre-sector region. A rather low velocity ratio in the post-sector is consistent with the asymmetric shape of the magnetic signature which shows a more pronounced B-field decrease at the inner side of the core / post-sector boundary. This sharp decrease suggests a small-scale current layer, possibly formed by hot expanding plasma of the core region, incorporated into the main current sheet.  The embedded current system affects predominantly the tangential magnetic field component and leaves $B_N$ unchanged. It is also associated with a jump in the motional electric field direction, with a ``toward'' orientation on both sides. Based on these signatures, the trailing edge of event 10 should provide a particularly efficient heating environment. The remote location of event 10 relative to the bow shock ($d_{BS} \sim 1.9 R_M$) could explain its intermittent nature.

\subsection{Non-HFA active current sheets}

\begin{figure*}
\noindent\includegraphics*[width=9 cm]{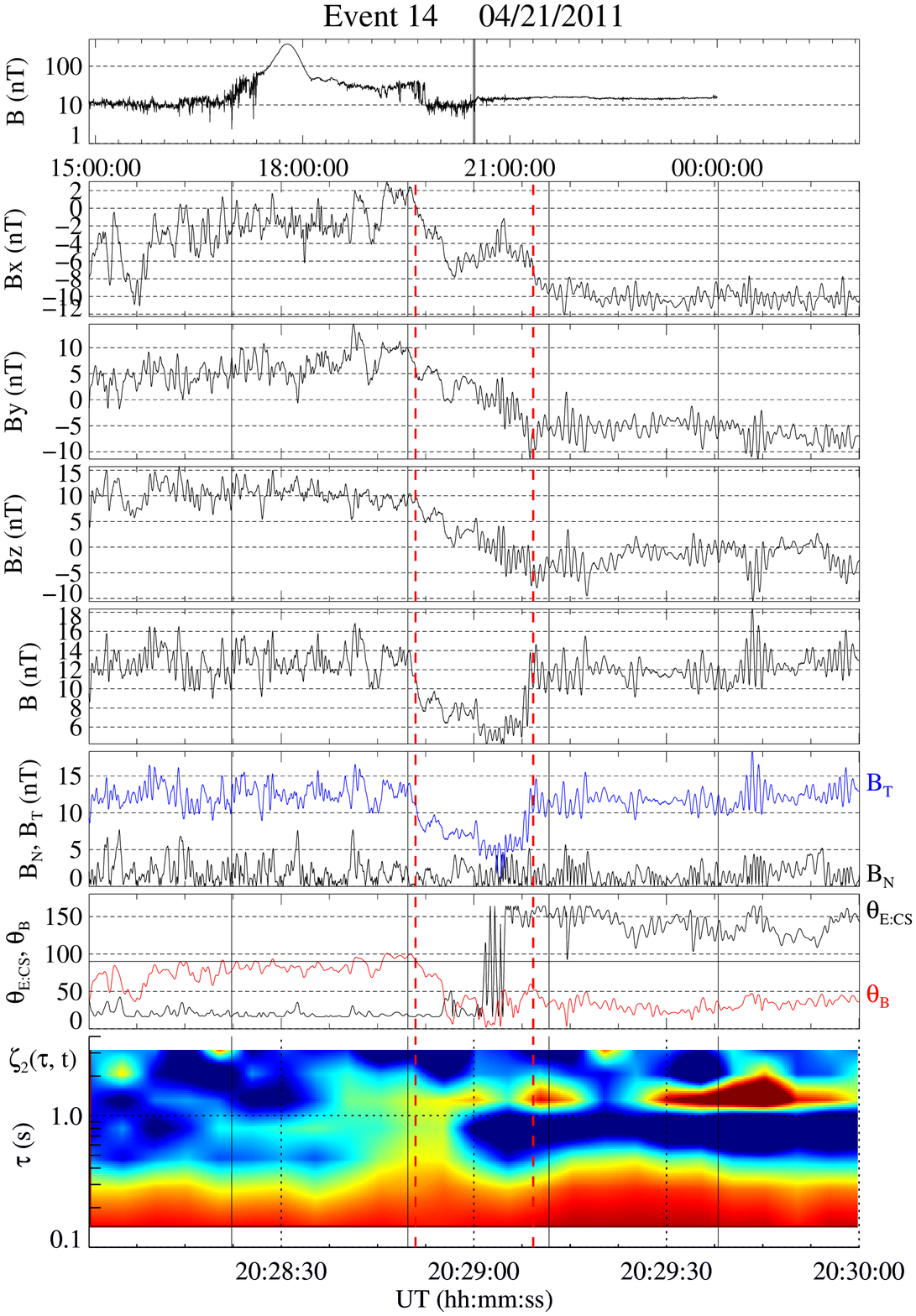}\noindent\includegraphics*[width=9 cm]{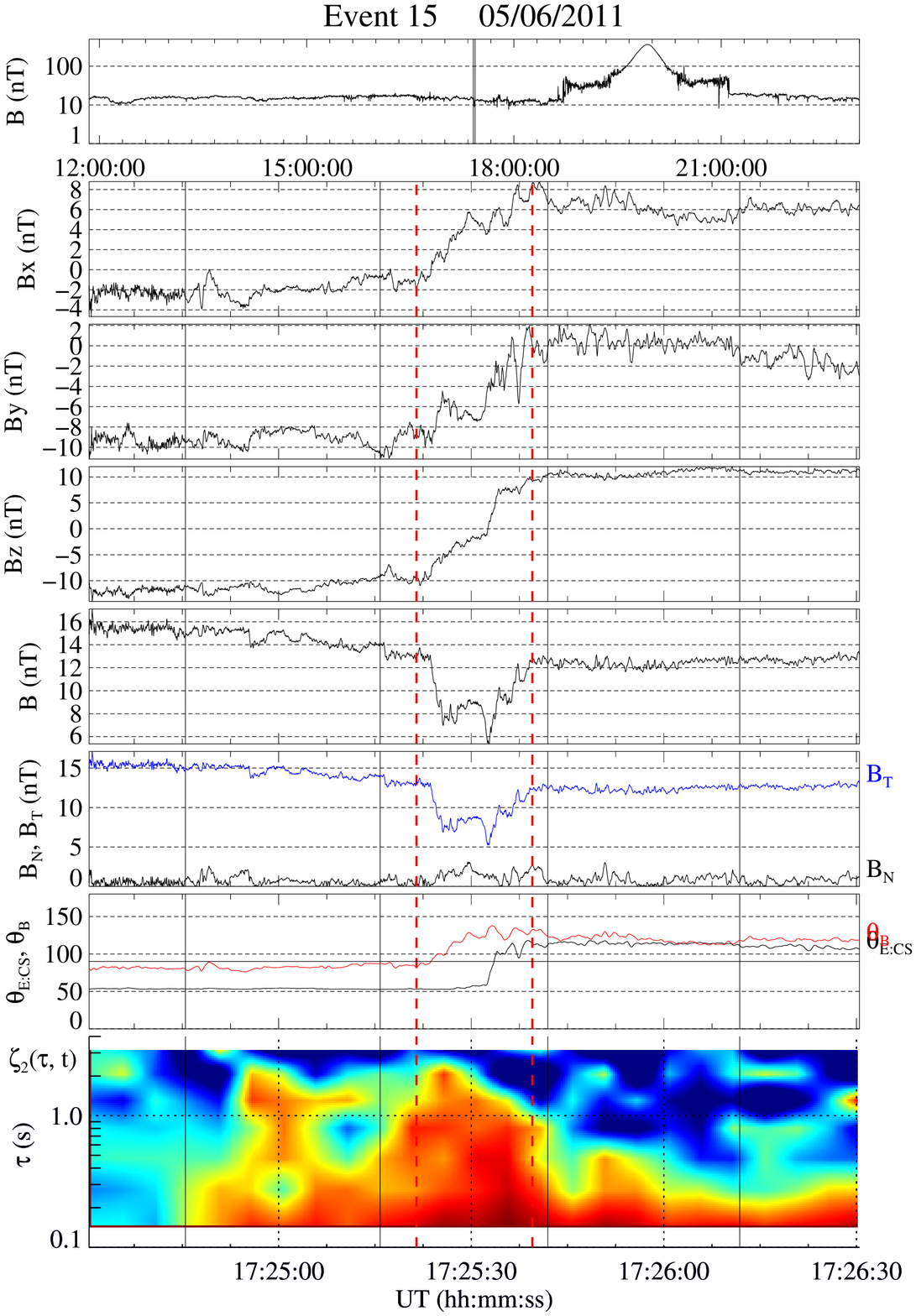}
\caption{\label{fig9} Examples of heliospheric current sheets in an upstream foreshock region.}
\end{figure*}
\begin{figure*}
\noindent\includegraphics*[width=9 cm]{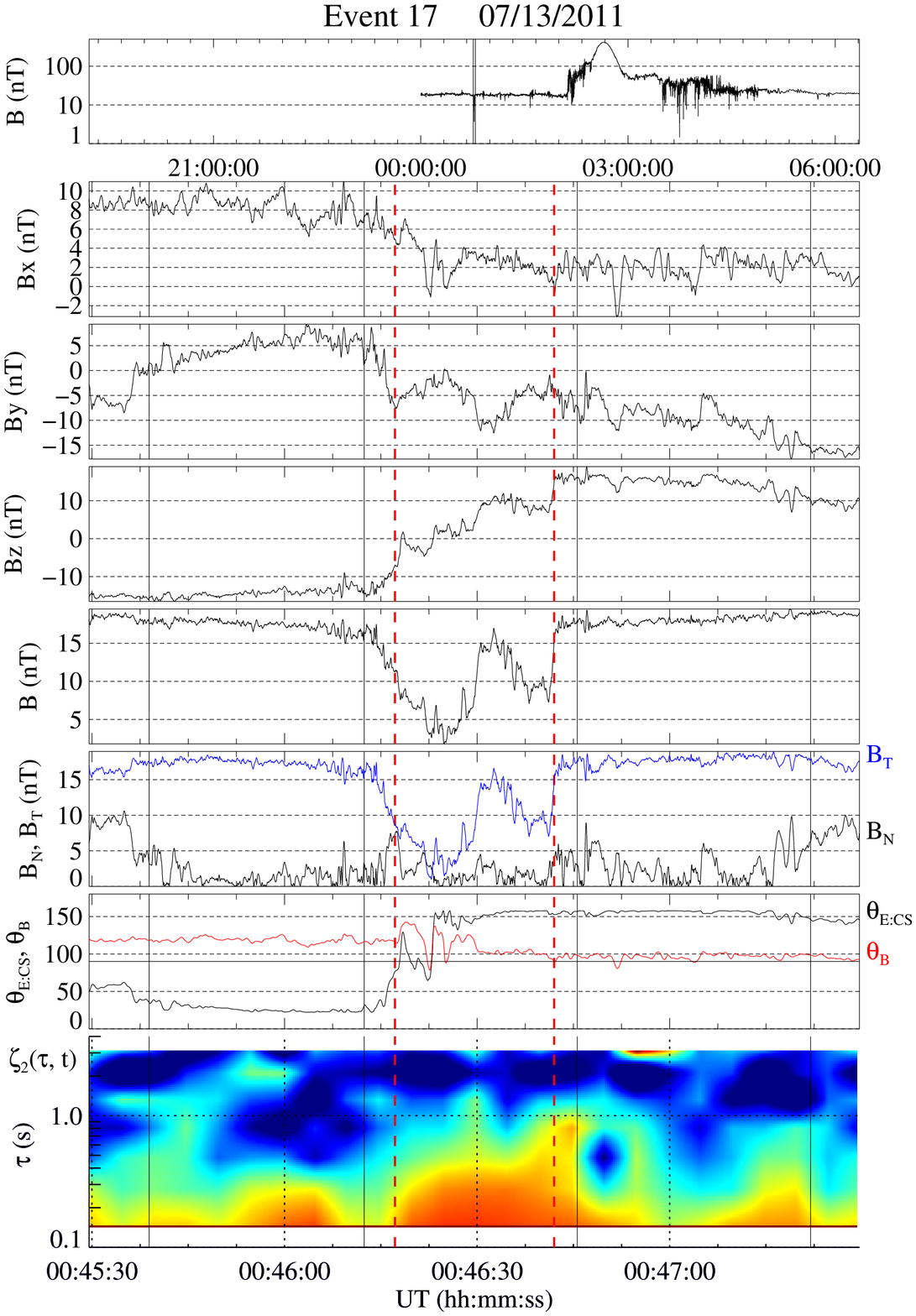}\noindent\includegraphics*[width=9 cm]{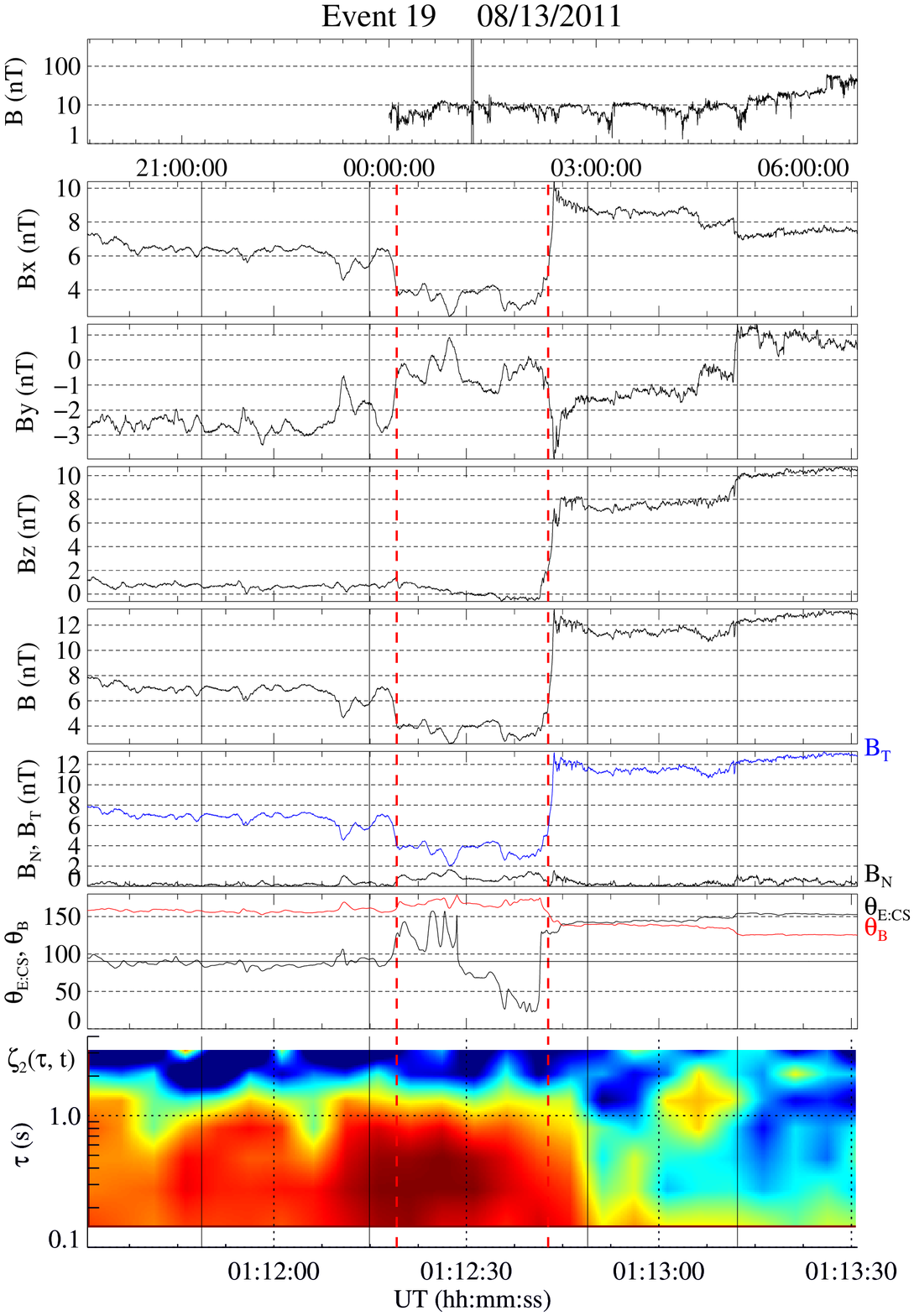}
\caption{\label{fig10} Examples of heliospheric current sheets in an upstream foreshock region.}
\end{figure*}

In addition to the HFA-like events described above, our algorithm has detected nine HCS events. Table \ref{table2} provides a summary of physical and geometric characteristics of these events. Four of them (Fig. \ref{fig9}-\ref{fig10}) are chosen to illustrate the distinction between HCS and HFA observations at Mercury's bow shock. These four non-HFA current sheets failed to develop HFA signatures by the time of their encounter with MESSENGER despite the proper (toward) orientation of the pre- and post-sector convection electric fields and a significant B-field depression in the middle. 

The HCS events shown in  Fig. \ref{fig9}-\ref{fig10} have no compressed leading or trailing edges characteristic of HFAs. They feature a large-scale reorganization of the field geometry across the current sheet. The directional E-field change across these current sheets is likely to represent a large-scale solar wind structure rather than a local plasma kinetics, and is more indicative of freely propagating helio current sheet than of  kinetically active current sheet interacting with the bow shock. All non-HFA helio current sheets were observed at a larger distance from the model bow shock compared to all but one HFAs on our list. The majority of the HCS events featured quasi-perpendicular magnetic field orientation relative to the local bow shock normal. These current sheets were likely to be magnetically disconnected from the bow shock, which prevented their evolution into HFAs. 

The fluctuation signatures of HCS events are also substantially different from those obtained for the HFA events. In helio current sheets, the increase of the ion kinetic crossover at the center of the sheet tends to be less dramatic compare to HFAs, and in some HCS events is completely missing. When the ion scale does increase, the enhancement was not well localized. 

{\bfseries Events 14 and 15} provide illustrative examples of such de-localized turbulent behavior. Event 14 detected on April 14, 2011 at $\sim$ 20:29:00 (Fig. \ref{fig9}, left) reveals the presence of ion-kinetic crossover (at $\tau_i \sim 0.3-0.4$ s) both long before and after the current sheet crossing, the type of behavior that HFA events usually do not show. There is a rather short transient increases of the ion scale value to about 1.2 s after which the scalogram returns to its background state. A more consistent increase of $\tau_i$ is observed during the passage of the trailing edge of event 14, with the upper range of ion kinetic scaling reaching 2 s. The blue-color coded gap which is present in the post-sector scalogram reveals a distorted shape of the structure function at the intermediate scales. The non-power law scaling associated with this gap could be a manifestation of a strong coherent oscillation inside the core region and the post-sector of the event. Polarization parameters of the oscillation are consistent with an obliquely propagating electromagnetic ion-cyclotron wave which could be excited by a two-stream instability caused by reflected solar wind ions. 

Event 15 observed on May 06, 2011 at around 17:25:30 (Fig. \ref{fig9}, right) demonstrates a minor ion heating (from $\sim 25$ to $\sim 45 $ eV) at the leading edge of the event, according to the shape of the SF scalogram. A similar transient temperature increase took place in the middle of the pre-sector of this HCS event showing that the heating was not limited to the interior of the current sheet. The $\tau_i$ enhancement during this event is more evident than that during event 14, but its spatial domain is poorly defined. 

{\bfseries Event 17} (Fig. \ref{fig10}, left) was detected on July 13 2011 at about 00:46:30. Two distinct magnetic depression regions possibly associated with embedded current sheets have been recorded during this event. Spatial orientation of the magnetic field remained approximately constant before and after the event. During the passage of the core region, the B-field  rotation angle reaches $\sim$ 140 degrees signaling strong directional discontinuity. The SF scalogram exhibits an increase of the ion kinetic time scale which is relatively well localized, with $\tau_i$ reaching its highest value ($\sim 0.8 - 1.0$ s) by the end of the event. 

{\bfseries Event 19} (Fig. \ref{fig10}, right) was encountered at about 00:46:30 on August 13 2011. The scalogram of this event reveals a rather hot plasma medium both in the core region and in the pre-sector for the event. This event shows a substantial change in the B-field direction and a fairly stable magnetic field orientation in the surrounding spatial domain. 

The trailing boundaries of the core regions of events 17 and 19 carry significant plasma discontinuities associated with abrupt changes of the magnetic intensity by about 7-8 nT over a  time interval 0.8-1.0 s. These B-field jumps correspond to a distance scale of the order of $400$ km, assuming $V_{SW}=450$ km/s. The proton Larmor radius characterizing these structures is about 80 km according to the eq. \ref{eq10} estimate. The discussed discontinuities are likely to have a kinetic origin while the large-scale HCS configuration seems to be formed in the MHD domain.

The rest of the HCS events identified by our code (see Table \ref{table2}) feature similar sets of signatures. Compared to HFAs, these events tend to occur at a more significant upstream distance from the nominal bow shock, with $d_B$ varying from 1.2 to 4.5 $R_M$). They feature a larger B-field rotation angle accompanied by a smaller relative change of the field magnitude indicative of rotational discontinuities (see \citet{schwartz00} and references therein), a weaker magnetic connection with the bow shock as suggested by the local shock geometry which is often quasi-perpendicular (especially after the events), and a considerably more stable B-field orientation in the pre- and post-sectors. The SF scalogram analysis of the HCS events reveals broader spatial regions occupied by the heated plasma contrasting with more compact and spatially localized heating during HFA events.

\section{Statistical survey}

\subsection{Event locations}

\begin{figure}
\begin{center}
\noindent\includegraphics*[width=12 cm]{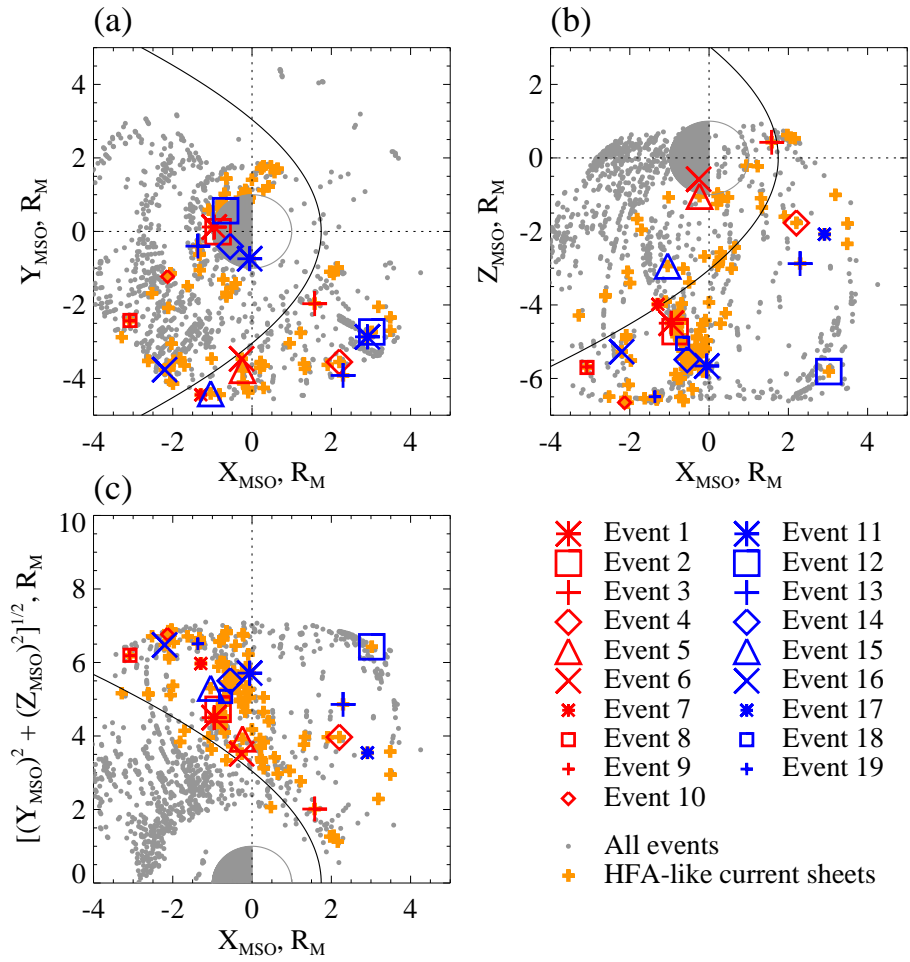}
\end{center}
\caption{\label{fig2} Occurrence locations of the detected events. See text for details.}
\end{figure}

Fig. \ref{fig2} presents MSO positions of all the detected events classified into several groups. Black dots ($n=1337$) show automatically detected magnetic depression events satisfying the condition (\ref{eq1a}) with $B_{th} = 5$nT. Yellow crosses ($n=100$) show filtered events which, in addition to the threshold condition, met a set of criteria making them candidate HFAs. These events had the correct (toward) orientation of the motional electric field on at least one side ($\theta_{E1:CS} \in [20, 70]$ and/or $\theta_{E2:CS} \in [110, 160]$), the duration $T$ of the core region between 10 and 100 s, and the distance $d_{BS}$ to the model bow shock lying between -0.5 and 3.0 $R_M$, with the negative (positive) $d_{BS}$ values corresponding to the event positions inside (outside) of the model bow shock boundary. 

Red symbols of different pattern designate the subset of the ten manually selected HFA-like events discussed in detail in the previous section. These events exhibit more reliable magnetic signatures of hot flow anomalies verified using a visual inspection and manual post-processing. The manually selected HFAs demonstrate identifiable compression edges on one or both sides of the event, a nearly perpendicular mutual orientation of $\mathbf{n}_{CS}$ and $\mathbf{n}_{BS}$ normals, and a consistently reduced B-field in the core region relative to the ambient pre- and post-sectors ensuring $\Delta B_0 > 0.5$. As discussed earlier in the text, the fluctuation signatures of these events are suggestive of an ongoing ion heating not expected for foreshock cavities.

Blue symbols are used for the set of nine HCS events discussed earlier in the text which mimic the geometry and large-scale magnetic signatures of the HFAs but lack the compressed edges and turbulent signatures of kinetically active current sheets providing local plasma heating. 

The first two planetary years of orbital MESSENGER operation covered a substantial portion of the dawn and dusk portions of the Hermean foreshock. The occurrence probability of all types of detected events is systematically higher in the dawn sector, which is in an agreement with terrestrial studies. It is known that on average, the dawn foreshock has a quasi-parallel magnetic field orientation allowing the reflected ions to be channeled onto the discontinuity. The post-midnight region characterized by large cone angles between the helio current sheets and the anti-sunward direction is of particular interest as it provides enough time for the kinetic processes to develop \citep{schwartz00}. A similar region at Earth is a preferred location of HFAs and a variety of other intermittent foreshock phenomena \citep{tsurutani85}. 

The automatically detected population of magnetic depression events at Mercury shown by black dots in Fig. \ref{fig2} may include a wide scope of activity not limited to HFA, in particular density holes -- subminute events sharing many properties of early stage HFAs \citep{parks06, wilber08}, foreshock cavities -- transient decreases of magnetic field strength and thermal ion density bounded by HFA-like edges of enhanced B-field \citep{billingham08}, and SLAMS growing out of the ULF wave field and featuring nonconvective electric fields at the edges \citep{schwartz92}. In its turn, the filtered subset of events (yellow crosses) may, in principle, contain a fraction of HFA or proto-HFA events although their HFA-like properties can not be established with certainty. The requirement of the inwardly directed motional electric field on one or both sides removes from our statistics most of the density holes which tend to show an opposite (outward) E-field orientation. The prolonged duration of the filtered events is not characteristic, albeit not impossible, for the SLAMS events observed in the terrestrial foreshock \citep{schwartz92}. 

As has been already noted, most of the manually validated HFA events (except for event 1) are located just outside of the nominal bow shock boundary (Fig. \ref{fig2}, see also Table \ref{table1}) while HCS events are observed systematically farther upstream. At Earth, hot flow anomalies are also formed when an interplanetary discontinuity with convergent motional electric field comes into direct contact with a quasi-parallel bow shock (within $\pm 20 \%$ from the model bow shock scaled by the observed solar wind dynamic pressure \citep{paschmann88}). HFAs passages are known to induce a significant magnetosheath response \citep{safrankova00, sibeck99} allowing the magnetopause to  move outward $\sim$ 5  planetary radii beyond its nominal position \citep{sibeck99} and causing significant effects in polar magnetosphere (see e.g. \citet{eastwood11}). Considering the proximity of the HFAs observed here to the bow shock boundary, one can expect that these events play an equally important role in Mercury's magnetosphere. 

\subsection{Event parameters}

\begin{figure}
\begin{center}
\noindent\includegraphics*[width=12 cm]{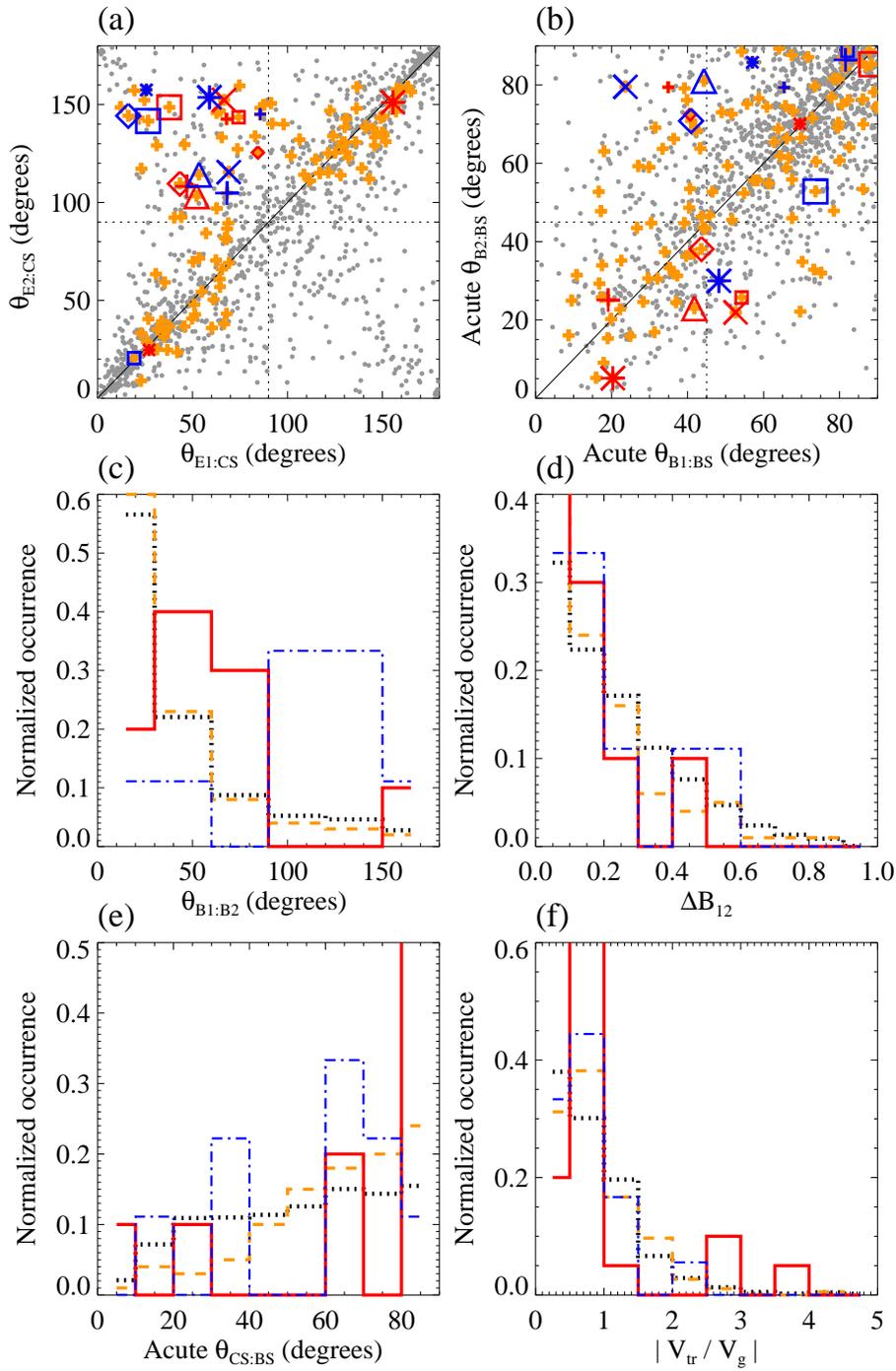}
\end{center}
\caption{\label{fig3} Statistical analysis of active current sheets in the Hermean foreshock; see Fig. \ref{fig2} for symbol notations.}
\end{figure}

Fig. \ref{fig3} presents several types of ensemble-based statistics of the detected events. 

Panel \ref{fig3}(a) shows the dependence between the pre-sector and post-sector E-field angle with the current sheet normal; panel \ref{fig3}(b) demonstrates a similar dependence for the bow shock normal angles with the magnetic field. Both panels use the same symbol notation as in Fig. \ref{fig2}. 

It can be seen that the majority of the unfiltered magnetic field depression events lie near the diagonal line $\theta_{E1:CS}=\theta_{E2:CS}$ corresponding to a constant motional electric field direction across the current sheet. For $\theta_{E1:CS} < 90^{\circ}$, the electric field is pointed toward the current sheet in the pre-sector, which causes the solar wind ions that are reflected by the bow shock immediately before the event to gyrate toward the current sheet. The condition $\theta_{E2:CS} > 90^{\circ}$ ensures similar behavior of the reflected particles at the trailing edge of the event. When the convection E-field points toward the discontinuity on both sides, the trajectories of the reflected ions from either side converge on the sheet and become channeled along it, see e.g. \citet{burgess89, schwartz95}. This configuration leading to the most efficient particle heating corresponds to the upper left corner of the plot. Eight out of ten events with definite HFA signatures are located in this quadrant. 

Panel Fig.\ref{fig3}(b) reveals no systematic tendency for the bow shock to become more quasi-perpendicular after the passage of automatically detected active current sheets. This tendency characterizes the behavior of Earth's bow shock before and after HFA events \citep{schwartz00}. Most of the HFA events validated manually, most noticeably events 1, 5, 6 and 8, seem to violate this rule making the trailing edge bow shock more quasi-parallel. It remains to be understood whether these exceptions represent a distinct physical tendency characteristic of HFAs at Mercury, or are caused by a statistical uncertainty.

The probability distributions of the angles between the pre- and post-sector magnetic field shown in Fig. \ref{fig3} (c) demonstrate significant differences. The automatically detected events, filtered and non-filtered, have a strong peak near $\theta_{B1:B2} \sim 0$ consistent with the fact that motional electric field remains unchanged across most of these events (since $\mathbf{V}_{SW}$ is kept constant). The manually selected HFA events show a systematically stronger B-field rotation peaking at 30-90 degrees which is in a good agreement with the statistics of terrestrial HFAs showing a maximum occurrence rate at $40-90^{\circ}$ \citep{schwartz00}. This is also consistent with the scattering the discontinuity angles describing interplanetary tangential discontinuities found at 0.46 to 0.5 AU \citep{lepping86}. The population of HCS events shows an even stronger magnetic rotation with the maximum occurrence rate at 90 - 150 degrees.

The distribution of the normalized change $\Delta B_{12}$ of the field magnitude across the current sheets presented in Fig. \ref{fig3}(e) provides some additional insight into the nature of the underlying interplanetary discontinuities. For a ``clean'' TD, $\Delta B_{12}$ is expected to be greater than 0.2, while smaller values correspond to mixed cases of either rotational or tangential discontinuity, depending on the normal magnetic field component. The $\Delta B_{12}$ histogram makes it clear that more than a half of the automatically detected HFA-like events belong to the second (mixed) category. A similar fraction of hot flow anomalies at Earth \citep{schwartz00} have this property. Considering that over $60 \%$ of the ``mixed'' cases may in fact also represent TDs \citep{neugebauer84}, the total number of tangential discontinuities in our database should dominate. This is important because our calculations of the current sheet normal is based on the assumption that the current sheets are TDs with a near-zero normal magnetic field component. This assumption is broadly used in single-spacecraft HFA studies. A more accurate identification of $\mathbf{n}_{BS}$ would require a minimum variance analysis through the current sheet which cannot be implemented in the presence of the embedded HFAs. 

The probability distribution of $\theta_{CS:BS}$ angles is biased toward large acute angles corresponding to perpendicular orientation of the current sheets relative to the local bow shock. It is similar to the corresponding distribution of terrestrial HFA events (compare with Fig. 10 of \citet{schwartz00}). The $\theta_{CS:BS}$ angle needs to be close to $90^{\circ}$ in order to keep transit velocity $V_{tr}$ sufficiently small to promote HFA formation. At Earth, about $80\%$ of the HFAs have acute $\theta_{CS:BS}$ angles greater than $60^{\circ}$. Our analysis also yeilds a significant fraction ($\sim 65\%$) of such events at Mercury. These events had an appropriate spatial orientation with respect to the bow shock which could contribute to their successful development into HFAs. This scenario is supported by the shape of the histogram of the normalized transit velocity $\left| V_{tr} / V_g \right|$ which was constructed by putting together pre- and post-sector measurements  (Fig. \ref{fig3}(f)). The histogram shows a prevailing occurrence rate (above $70\%$) of smaller-than-one velocity ratios, even though most of the events were observed away from the subsolar point providing optimal current sheet - bow shock intersection conditions. The manually validated HFA events have the highest occurrence frequency of smaller-than-unity velocity ratios among the studied groups of events.

\begin{table} \caption{Parameters of HFA-like events}\label{table1}
\begin{tabular}{lcccccccccc}
\hline
Event & 1 &  2 &  3 &  4 &  5 &  6 &  7 & 8 &  9 & 10 \\
\hline
mm/dd & 04/16 &  04/16 &  04/20 &  04/21 &  05/04 &  05/04 &  05/07 &  08/13 & 
08/13 &  08/13 \\
$\frac{(t'_0\,\,+\,\,t''_0)}{2}$ & 19:00:20 &  19:09:10 &  04:50:44 &  03:52:52 &  06:08:30 &  
18:26:42 &  04:46:02 &  04:28:04 &  04:29:12 &  02:22:48 \\   
\hline
$X_{MS0}$  &  -0.96 &   -0.86 &    1.58 &    2.20 &   -0.24 &   -0.26 & 
 -1.29 &   -3.08 &   -3.09 &   -2.13 \\
$Y_{MS0}$  &   0.12 &   -0.01 &   -1.97 &   -3.55 &   -3.78 &   -3.47 & 
 -4.45 &   -2.42 &   -2.43 &   -1.23 \\
$Z_{MS0}$  &  -4.50 &   -4.73 &    0.42 &   -1.77 &   -1.04 &   -0.58 & 
 -3.98 &   -5.70 &   -5.69 &   -6.65 \\
$d_{BS}$   &  -0.56 &   -0.80 &   -0.50 &   -2.21 &   -0.52 &   -0.19 & 
 -1.60 &   -0.90 &   -0.88 &   -1.86 \\
\hline
$\theta_{B1:B2}$  &   18 &   158 &    44 &    81 &    63 &    73 &     2 & 
  36 &    50 &    32 \\
$\theta_{E1:CS}$  &  155 &    37 &    46 &    43 &    52 &    66 &    27 & 
  74 &    67 &    84 \\
$\theta_{E2:CS}$  &  151 &   148 &   108 &   109 &   103 &   152 &    24 & 
 143 &   142 &   125 \\
$\theta_{B1:BS}$  &  159 &    88 &   160 &   136 &    41 &    52 &    69 & 
  54 &    34 &    40 \\
$\theta_{B2:BS}$  &  174 &    94 &   154 &   141 &    22 &    21 &    70 & 
  25 &    79 &    72 \\
$\theta_{CS:BS}$  &   94 &     8 &    90 &    88 &    96 &    96 &    24 & 
 111 &   112 &    99 \\
$\theta_{SW:CS}$  &  112 &   119 &   120 &   130 &   134 &   114 &    98 & 
 110 &   115 &   123 \\
\hline
$\Delta B_{12}$ &   0.20 &    0.05 &    0.41 &    0.13 &    0.08 &    0.09 & 
  0.15 &    0.08 &    0.15 &    0.09 \\
$\Delta B_{0}$  &   1.09 &    0.77 &    2.06 &    0.48 &    1.02 &    1.07 & 
  0.62 &    0.37 &    0.37 &    0.28 \\
$\Delta t$, s       &   20 &    21 &    18 &    12 &    10 &     6 &   287 & 
  17 &    13 &     4 \\
\hline
$| V_{tr}/V_g |_1 $ &   0.98 &    2.87 &    0.92 &    0.64 &    0.86 & 
  0.42 &    0.35 &    0.48 &    0.86 &    0.88 \\
$| V_{tr}/V_g |_2 $ &   3.83 &    2.88 &    0.71 &    0.71 &    1.46 & 
  0.89 &    0.35 &    0.90 &    0.50 &    0.60 \\
\hline
\hline
\end{tabular}
\end{table}
\begin{table} \caption{Parameters of HCS events}\label{table2}
\begin{tabular}{lccccccccc}
\hline
Event & 11 &  12 &  13 &  14 &  15 &  16 &  17 & 18 &  19 \\
\hline
mm/dd & 04/11 &  04/13 &  04/21 &  04/21 &  05/06 &  05/11 &  07/13 &  07/27 & 
08/13 \\
$\frac{(t'_0\,\,+\,\,t''_0)}{2}$ & 19:12:18 &  11:30:20 &  03:13:34 &  20:29:00 &  17:25:30 & 
04:11:09 &  00:46:29 &  23:24:28 &  01:12:30 \\
\hline
$X_{MS0}$  &  -0.06 &    3.02 &    2.30 &   -0.55 &   -1.04 &   -2.20 & 
  2.91 &   -0.67 &   -1.37 \\
$Y_{MS0}$  &  -0.74 &   -2.74 &   -3.92 &   -0.40 &   -4.41 &   -3.76 & 
 -2.87 &    0.56 &   -0.40 \\
$Z_{MS0}$  &  -5.66 &   -5.81 &   -2.88 &   -5.49 &   -2.94 &   -5.27 & 
 -2.08 &   -5.04 &   -6.50 \\
$d_{BS}$   &  -2.07 &   -4.54 &   -2.89 &   -1.62 &   -1.18 &   -1.57 & 
 -2.47 &   -1.20 &   -2.02 \\
\hline
$\theta_{B1:B2}$  &  116 &    97 &   170 &   108 &   139 &   123 &   141 & 
  23 &    38 \\
$\theta_{E1:CS}$  &   58 &    26 &    68 &    16 &    53 &    69 &    25 & 
  19 &    85 \\
$\theta_{E2:CS}$  &  153 &   141 &   104 &   144 &   114 &   115 &   157 & 
  20 &   145 \\
$\theta_{B1:BS}$  &  131 &   106 &    81 &   139 &    44 &    23 &    57 & 
  97 &    65 \\
$\theta_{B2:BS}$  &   29 &    52 &    93 &   109 &    98 &   100 &    85 & 
  91 &   100 \\
$\theta_{CS:BS}$  &   67 &    39 &    31 &   105 &    67 &    83 &   105 & 
  17 &    65 \\
$\theta_{SW:CS}$  &  115 &   114 &   138 &   106 &   142 &   153 &   112 & 
 109 &   112 \\
\hline
$\Delta B_{12}$ &   0.15 &    0.13 &    0.52 &    0.09 &    0.15 &    0.26 & 
  0.03 &    0.10 &    0.42 \\
$\Delta B_{0}$  &   0.48 &    0.21 &    0.88 &    0.70 &    0.57 &    0.36 & 
  0.98 &    0.24 &    0.54 \\
$\Delta t$, s       &   14 &    31 &    10 &    18 &    18 &     9 &    24 & 
   3 &    23 \\
\hline
$| V_{tr}/V_g |_1 $ &   0.56 &    0.50 &    1.03 &    0.39 &    1.12 & 
  2.25 &    0.30 &    0.99 &    0.45 \\
$| V_{tr}/V_g |_2 $ &   0.83 &    0.60 &    1.02 &    0.27 &    0.79 & 
  0.92 &    0.25 &    0.98 &    0.42 \\
\hline
\hline
\end{tabular}
\end{table}





\subsection{Hermean HFA sizes compared to other planets}


It presents a certain interest to compare the estimated sizes of typical and extreme HFA events at Mercury with the corresponding sizes of HFA events reported for Earth and Saturn. 

To derive characteristic linear sizes of Hermean HFA events, we multiplied the duration of the core region of each event by the nominal solar wind velocity of 450 km/s, and applied the correction factor $\cos (\theta_{SW:CS})$ accounting for the $\mathbf{n}_{CS}$ misalignment with the solar wind flow. Event 7 whose duration was by an order of magnitude longer that the duration of other events was excluded from the statistics and treated separately. 

The typical size of terrestrial HFAs were taken from a global statistical survey conducted by \citet{facsko09}. Two of their estimates were used: the one based on the Alfv\'{e}n speed (method 1) and the one based on the speed of the TD and bow shock intersection calculated from {\it in situ} solar wind measurements (method 2). 

As an example of an extreme terrestrial HFA we choose an event reported by \citet{safrankova12} based on multi-spacecraft observations. This event had a rather significant size and caused a considerable magnetosphere deformation. 

The typical size of HFA events at Saturn was evaluated using the data reported by \citet{masters09}; the largest event described in their paper was considered as an extreme event. The Kronian HFA sizes were evaluated for a nominal solar wind velocity of 450 km/s taking into account spatial orientation of the current sheets.

\begin{figure*}
\noindent\includegraphics*[width=11 cm]{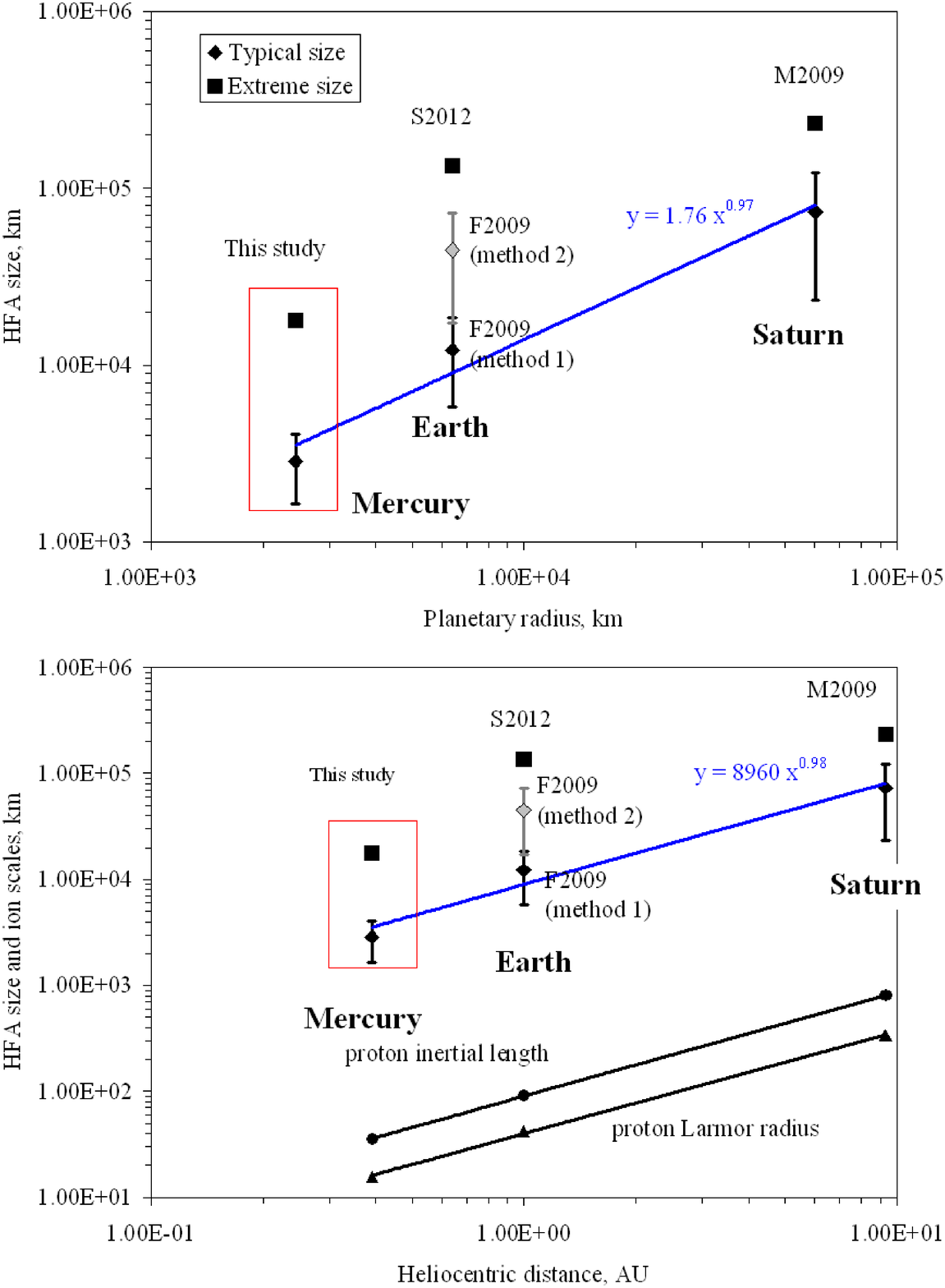}

\caption{\label{fig11} Estimated linear sizes of HFA events at Mercury as compared to the sizes of the HFA events at Earth and Saturn, as a function of planetary radius (top) and heliocentric distance (bottom). The typical sizes of terrestrial HFAs are adopted from \citet{facsko09} who used two complementary techniques (labeled as ``method 1'' and ``method 2'') to get identify HFAs. The size of an extreme terrestrial event is taken from \citet{safrankova12}. The case studies published by \citet{masters09} were used to evaluate typical and extreme event sizes of HFA events at Saturn. The solid blue lines are the best power law fits to the data (excluding the method 2 - based estimate for terrestrial events). The ion scale plots are constructed using the statistical solar wind model by \citet{kohnlein96}. The data suggest that the typical characteristic size of planetary HFAs is approximately 1.2 planetary radii which could be a combined effect of the bow shock geometry and the increase of the proton scales with the distance from the Sun. }
\end{figure*}

Fig. \ref{fig11} presents the comparison of the event sizes at the three planets. 

As can be seen, the size of the typical HFA tends to increase with the size of the planet. The dependence is approximately linear. To the first approximation, the size of the HFA at Mercury, Earth and Saturn is of the order of one planetary radius. The observed dependence can indicate that spatial dimensions of HFAs are controlled by the geometry of the bow shock, with the largest (Kronian) events formed at the least curved bow shock boundary. 

The data also show that the HFA size is directly proportional to the heliocentric distance (Fig. \ref{fig11}, bottom panel). This is an expected behavior since the sizes of the three compared planets scale approximately linearly with the distance from the Sun. The increase of the ion scales describing an expanding solar wind can contribute to the observed dependence. To illustrate this possibility, the figure provides characteristic values of proton inertial length and proton gyro radius calculated using the data-derived statistical solar wind model developed by \citep{kohnlein96}. It can be seen that both kinetic scales grow linearly with the heliocentric distance in proportion to the HFA size.

\section{Concluding remarks}

We have presented first documented observations of HFA-like events at Mercury. Using magnetic and particle data from MESSENGER collected over a course of two planetary years, we identified a representative ensemble of active current sheets magnetically connected to Mercury's bow shock. Some of these events exhibit unambiguous magnetic and particle signatures of hot flow anomalies. A broader subset of the detected magnetic depression events is likely to include a variety of plasma disturbances not limited to HFAs, such as the disturbances associated with density holes, foreshock cavities, and SLAMS. 

Our classification of current sheets as HFAs is based on an investigation of the current sheet geometry involving all of the commonly used aspects such as the direction of the motional electric field, the bow shock location, the orientation of the current sheet and bow shock normals relative to the solar wind flow, and on a number of additional tests. Four of the reported HFA events showed unambiguous signatures of ion energization documented by FIPS dynamic spectra confirming the presence of heated plasma inside and around the current sheets. Although accurate FIPS measurements have limited time resolution and angular coverage, they provided a key piece of evidence by revealing hot plasma populations at the expected locations.

HFA events at Mercury are accompanied by a systematic change in the magnetic turbulence spectrum predicted for a locally energized solar wind plasma. Our previous study \citep{uritsky11} has shown that the Hermean magnetosphere, as well as the surrounding region, are affected by non-MHD effects introduced by finite sizes of cyclotron orbits of the constituting ion species. These results have demonstrated that plasma fluctuations at this planet are largely controlled by finite Larmor radius effects. The heating process associated with HFA-like active current sheets explored in the present paper are an important manifestation of such behavior at Mercury.

One one occasion, we detected signatures of a ULF wave packet in a quasi-parallel shock configuration which was likely to be triggered by an HFA event. Such upstream large-amplitude waves may propagate deep into Mercury's magnetosphere causing secondary instabilities in various plasma regions. They can also reach the surface through the thin atmosphere not protected by a conducting ionosphere. 

The occurrence rate of HFA events at Mercury is systematically higher in the dawn sector compared to the dusk sector. On average, the dawn foreshock has a quasi-parallel magnetic field orientation allowing the reflected ions to be channeled onto the discontinuity. A similar region at Earth is a preferred location of HFAs and a variety of other intermittent foreshock phenomena \citep{tsurutani85}. The post-midnight region characterized by large cone angles between the helio current sheets and the anti-sunward direction is of particular interest as it provides enough time for the kinetic processes to develop. Most of our manually validated HFA events were encountered just outside of the nominal bow shock boundary. Because of their proximity to the bow shock boundary, such events can play an important role in Mercury's magnetosphere. Terrestrial HFAs passages are known to induce a significant magnetosheath response \citep{safrankova00, sibeck99} allowing the magnetopause to move outward $\sim$ 5 planetary radii beyond its nominal position and perturbing the magnetosphere (see e.g. \citet{eastwood11}). It would be important to verify in future studues whether similar global HFA-induced  phenomena occur at Mercury. As an indirect manifestation of such behavior, there is a slight tendency for the bow shock to become more quasi-perpendicular after the passage of HFA-like events, which is also in line with the behavior of Earth's bow shock before and after HFAs \citep{schwartz00}.

The characteristic linear size of HFA events at Mercury is noticeably smaller than that on other planets. When combined with previously reported HFA sizes at Earth and at Saturn, our measurements show that the size of planetary HFAs grows linearly with planetary radius. The observed dependence could reflect the bow shock geometry. An alternative explanation takes into account the fact that the size of the compared planets is proportional to their distance from the Sun. The increase in the heliocentric distance leads to larger ion scales in the expanding solar wind controlling the thickness of helio current sheets, and hence the size of the HFAs.

In summary, we have demonstrated that Mercury's bow shock contains HFA events similar to those observed at other planets. The conducted quantitative analysis suggests that Mercury's bow shock provides conditions for local particle acceleration and heating as predicted by previous numerical simulations. Together with earlier observations of HFA activity at Earth, Venus and Saturn,  our results confirm that hot flow anomalies are a common property of planetary bow shocks.









\begin{acknowledgments}
The work of V.U. was supported by the NASA grant NNG11PL10A 670.002 through the CUA's Institute for Astrophysics and
Computational Sciences.
\end{acknowledgments}




\newpage

\bibliographystyle{agu04}
\bibliography{grl_2010} 


\end{article}

\end{document}